{
\documentclass{aa}

\usepackage{graphicx}
\usepackage{txfonts}
\newcommand{\kms}{km~s$^{-1}$\,}
\newcommand{\msun}{${\cal M}_\odot$\,}

\begin{document}

   \title{Spectroscopic orbits of nearby stars}

   \author{J. Sperauskas
          \inst{1} \and
          V. Deveikis \inst{1} \and
          A. Tokovinin \inst{2}
           }

\institute{Vilnius University Observatory, Saul{\.e}tekio al. 3, 10257 Vilnius, Lithuania\\
              \email{julius.sperauskas@ff.vu.lt}
         \and
Cerro Tololo Inter-American Observatory, Casilla 603, La Serena, Chile \\
\email{atokovinin@ctio.noao.edu}
             }

   \date{Received  2019; accepted }


  \abstract
   {}
   {We observed stars with variable radial velocities  to determine their spectroscopic orbits.}
   {Velocities of  132 targets taken over a time span reaching 30 years
are presented. They were measured with the correlation radial velocity spectrometers (1913 velocities) and the new
VUES echelle spectrograph (632 velocities), with typical accuracy of 0.5 and 0.2 \kms, respectively. }
   {We derived spectroscopic orbits of 57
   stars   (including 53 first-time orbits),  mostly nearby dwarfs
   of spectral types K and M.  Their periods range from 2.2 days to 14
   years, some of those  are Hipparcos astrometric binaries.  Comments
   on  individual   objects  are  provided.   Many   stars  belong  to
   hierarchical systems containing three or more components, including
   20 new  hierarchies resulting  from this project.   The preliminary
   orbit  of  the  young  star  HIP~47110B has  a  large  eccentricity
   $e=0.47$  despite  short  period  of  4.4  d;  it  could  be  still
   circularizing.}
{Our   results  enrich the  data on  nearby stars and  contribute  to a better
   definition of the multiplicity statistics.}


   \maketitle
%

\section{Introduction}
\label{sec:intro}

Solar neighbourhood is the best studied part of the Galaxy. Yet, it is
still a site  of active research and new  discoveries.  While the {\it
  Gaia} satellite  is improving the census of  our stellar neighbours,
several  observational  campaigns target  nearby  stars  in search  of
exo-planets. Studies  of specific  stellar populations, such  as young
associations  or metal-poor  stars,  inevitably focus  on the  nearest
objects. Last, but not least,  solar neighbourhood is the benchmark for
stellar multiplicity statistics.

Here we  report the results of  the large campaign  of radial velocity
(RV) measurements  targeting mostly nearby low-mass  stars. It started
three decades ago,  before the  {\it Hipparcos} mission.  The aims  and main
results of this campaign are presented by \citet{S16}, where the stars
with constant RVs are featured. Here we focus on the remaining objects
with  variable   (or  supposedly  variable)  RVs.  Our   goal  is  the
determination of spectroscopic orbits of these stars.

RV monitoring  during several years  is very efficient  in discovering
spectroscopic binaries (SBs); only a few (e.g.  three) RV measurements
suffice to  detect the  RV variability or  double lines.  However,
determination of orbits of  these SBs requires  substantial follow-up
efforts.  For  example, the Geneva-Copenhagen  Survey, GCS \citep{N04},
discovered  hundreds of  SBs, but  their orbits  remain, for  the most
part,  unknown or  unpublished.  This  leads to  uncertainties  in the
study of  stellar multiplicity, namely in the  distribution of periods
and mass ratios of nearby solar-type stars \citep{FG67}.

Nowadays,  hundreds of thousands  of RV  measurements are  coming from
{\it  Gaia}  \citep{Gaia} and  ground-based  surveys  such as  APOGEE
\citep[e.g.][]{APOGEE},      RAVE     \citep{RAVE},      or     LAMOST
\citep{LAMOST}.  However, some SB  orbits   require either  a long
time  span or  a  frequent  cadence, not  furnished  by the  automatic
surveys.  So far, only {\it Gaia} offers the full-sky coverage, but it
has not  yet provided individual  RVs measurements suitable  for orbit
calculation. Moreover,  treatment of double-lined  systems and complex
cases such as  triples by the {\it Gaia}  pipeline may be problematic.
Our  results,  therefore,  are  unlikely  to become  obsolete  in  the
near term.

The  results of our  observations will  be useful  in many  ways. They
provide previously  unknown periods and mass ratios  of low-mass binaries
in the solar neighbourhood that  serve to improve the multiplicity
statistics. Recent discovery of  the strong dependence of close-binary
fraction  on metallicity \citep{Moe2018}  puts such  efforts in  a new
context.  Several  binaries detected by  {\it Hipparcos} accelerations
have their spectroscopic orbits determined here.  Similarly, this data
will help in the  interpretation of astrometric accelerations measured
by {\it Gaia},  as our time coverage is much  longer than the duration
of this mission.   Finally, some objects in our  sample present special
interest for various reasons,  making essential the knowledge of their
orbits.

The objects of this study are presented in Sect. 2. Sect. 3 covers
the instruments and methods used to derive the orbits. Our results,
namely the orbits and comments on individual objects, are given in
Sect. 4. The conclusions in Sect. 5 close the paper.

\section{The observed sample}
\label{sec:sample}

\longtab{
\begin{longtable}{l cc  rr ll rr r rrc}
\caption{Object list  \label{tab:list}  }  \\
\hline\hline
Name & RA & Dec & $V$ & $\varpi$ & Sp.  & Type & $N$ & $\Delta T$ &
$\langle$RV$\rangle$ & rms & $\chi^2/(N-1)$ \\
 &  \multicolumn{2}{c}{J2000} & (mag) & (mas) &  type & && (d) & (km~s$^{-1}$) & (km~s$^{-1}$) & \\
\hline
\endfirsthead
\caption{continued.}\\
\hline\hline
Name & RA & Dec & $V$ & $\varpi$ & Sp. & Type & $N$ & $\Delta T$ &
$\langle$RV$\rangle$ & rms & $\chi^2/(N-1)$ \\
 &  \multicolumn{2}{c}{J2000} & (mag) & (mas) &  type & && (d) & (km~s$^{-1}$) & (km~s$^{-1}$) & \\
\hline
\endhead
\hline
\endfoot
HIP 96       &  00 01 13.19 &  $+$13 58 30.3 & 10.59 &   23.25 & M0.5  & V  &   7 &   2555 &    $-$7.93 &   1.49 &   23.6  \\
BD+13 5195B  &  00 01 12.87 &  $+$13 58 19.7 & 11.12 &   27.77 & M1    & V  &   5 &   2541 &   $-$10.15 &   0.82 &   10.4  \\
HIP 374      &  00 04 40.08 &  $+$34 15 54.4 &  7.08 &    5.64 & K0    & C  &   4 &   6231 &   $-$24.02 &   0.35 &    1.8  \\
HIP 375      &  00 04 40.19 &  $+$34 16 19.8 & 10.13 &    1.00 & --    &  C &   4 &   6242 &    $-$9.17 &   0.24 &    0.9  \\
BD+33 4827D  &  00 04 33.54 &  $+$34 15 03.5 & 10.55 &    4.56 & F9    & S1 &  49 &   6242 &   $-$23.48 & \ldots & \ldots  \\
HIP 1412     &  00 17 40.90 &  $-$08 40 56.2 & 10.95 &   31.57 & K7V   & V  &   6 &   1479 &    15.73 &  14.41 &   99.0  \\
HIP 3428     &  00 43 41.42 &  $+$23 53 07.1 & 10.97 &   22.18 & K7    & S2 &  16 &   2168 &    $-$6.65 & \ldots & \ldots  \\
HIP 5110A    &  01 05 29.92 &  $+$15 23 24.2 &  9.17 &   36.65 & K3.5V & C  &   4 &   2915 &    $-$5.72 &   0.35 &    2.4  \\
HD 6440B     &  01 05 29.76 &  $+$15 23 15.5 &  9.93 &   37.41 & K8V   & V  &   8 &   2915 &    $-$3.22 &   1.23 &   12.7  \\
HD 8691      &  01 26 09.28 &  $+$31 54 52.6 &  9.24 &   19.96 & G0    & S1 &  42 &   5440 &   $-$36.18 & \ldots & \ldots  \\
HIP 9867     &  02 06 57.21 &  $+$45 11 04.1 & 10.24 &   51.47 & M0V   & S2 &  17 &   5479 &    61.42 & \ldots & \ldots  \\
HIP 10258    &  02 11 57.98 &  $+$04 21 41.8 &  9.72 &   21.36 & K5    & S2 &  30 &   4742 &   $-$59.15 & \ldots & \ldots  \\
BD+49 646    &  02 22 33.89 &  $+$50 33 36.9 &  9.65 &   18.91 & --    & s2 &   8 &   4711 &   $-$10.00 & \ldots & \ldots  \\
HIP 11437    &  02 27 29.25 &  $+$30 58 24.6 & 10.12 &   24.36 & K7V   & C  &  10 &   5884 &     5.98 &   0.46 &    1.8  \\
HIP 12787    &  02 44 21.36 &  $+$10 57 41.3 & 11.10 &   20.51 & M0Ve  & s2 &  12 &   2970 &     4.20 & \ldots & \ldots  \\
HIP 13398    &  02 52 25.03 &  $+$26 58 29.9 & 11.05 &   42.67 & M2V   & V? &  13 &   5459 &   $-$11.59 &   0.74 &    3.7  \\
HIP 13460    &  02 53 18.50 &  $+$60 51 11.7 &  9.20 &   25.75 & K3V   & S2 &  30 &   2965 &   $-$73.61 & \ldots & \ldots  \\
HIP 14478    &  03 06 51.34 &  $+$40 21 33.5 &  9.63 &   37.02 & K6    & S1 &  15 &   2917 &   $-$50.87 & \ldots & \ldots  \\
HIP 14669    &  03 09 30.79 &  $+$45 43 57.9 & 10.17 &   57.11 & M2V   & V  &   9 &   2629 &    $-$4.51 &   1.99 &   35.2  \\
HIP 14864    &  03 11 56.83 &  $+$61 31 13.0 & 10.05 &   39.82 & M0Ve  & S2 &  33 &   3023 &   $-$32.10 & \ldots & \ldots  \\
BD+03 480    &  03 28 14.92 &  $+$04 09 47.4 &  9.49 &   11.99 & G0    & C  &  11 &   4764 &    12.66 &   0.57 &    1.1  \\
HIP 17102    &  03 39 48.96 &  $+$33 28 24.3 &  9.05 &   25.49 & K2    & S2 &  23 &   2777 &     4.27 & \ldots & \ldots  \\
GJ 3248      &  03 48 38.24 &  $+$73 32 35.3 & 11.32 &   62.71 & M1V   & V? &  11 &   2971 &    $-$8.08 &   0.78 &    5.0  \\
HIP 18448    &  03 56 36.22 &  $+$69 50 55.9 &  9.33 &    6.71 & K0    & S1 &  29 &   6014 &    16.66 & \ldots & \ldots  \\
HIP 19410    &  04 09 26.37 &  $-$14 41 54.1 & 10.61 &   24.67 & K5V   & V  &   3 &   4350 &     6.34 &   7.97 &   99.0  \\
HIP 19915    &  04 16 19.83 &  $+$36 44 02.8 &  8.98 &    6.01 & F8    & S2 &  41 &  10496 &    $-$8.16 & \ldots & \ldots  \\
HD 279846    &  04 26 09.68 &  $+$34 09 34.2 & 10.50 &   12.14 & K2    & S2 &  28 &   6089 &    12.43 & \ldots & \ldots  \\
HIP 20709    &  04 26 15.05 &  $+$34 42 57.2 &  8.29 &    7.60 & F5    & S2 &  27 &   4743 &    44.76 & \ldots & \ldots  \\
HIP 21710    &  04 39 42.61 &  $+$09 52 19.5 &  9.19 &   36.80 & K2    & S1 &  16 &   1340 &   $-$25.93 & \ldots & \ldots  \\
HIP 21845    &  04 41 47.25 &  $+$28 39 35.9 &  8.85 &    8.70 & F8    & V? &  11 &   5191 &    11.20 &   0.59 &    2.3  \\
HIP 23550    &  05 03 51.96 &  $+$24 58 22.1 &  7.43 &   13.69 & G8V   & V  &  16 &   3271 &    54.18 &   1.18 &   10.6  \\
HIP 24488    &  05 15 15.46 &  $+$47 10 14.6 &  6.92 &    8.73 & G5III & s2 &  10 &   7146 &    24.42 & \ldots & \ldots  \\
GJ 220       &  05 53 14.04 &  $+$24 15 32.9 & 10.82 &   51.50 & M2.0V & V  &  13 &   6089 &    26.45 &   1.75 &   28.2  \\
HIP 28663    &  06 03 08.64 &  $+$14 21 54.4 &  8.31 &    9.74 & F4IV  & S1 &  30 &   1722 &   $-$25.10 & \ldots & \ldots  \\
HIP 29295    &  06 10 34.62 &  $-$21 51 52.7 &  8.12 &  173.70 & M1V   & C  &   6 &   5754 &     4.71 &   0.32 &    0.3  \\
HIP 29316    &  06 10 54.80 &  $+$10 19 05.0 & 10.39 &   91.65 & M2.5V & V  &  10 &   2567 &    52.81 &   1.27 &   15.4  \\
HIP 30269    &  06 22 02.50 &  $-$05 27 17.0 &  8.06 &    3.22 & F5V   & V  &  14 &   5389 &    20.29 &  14.85 &   99.0  \\
HIP 33560    &  06 58 26.05 &  $-$12 59 30.6 &  9.16 &   44.41 & K4V   & V? &  14 &   8376 &    $-$4.33 &   0.96 &    2.3  \\
HIP 34341    &  07 07 09.31 &  $+$03 26 50.7 &  9.87 &   38.26 & K5    & S1 &  21 &   5703 &   $-$17.35 & \ldots & \ldots  \\
HIP 35706    &  07 22 02.05 &  $+$68 16 27.6 & 10.10 &   23.84 & K5V   & S1 &  21 &   5824 &   $-$29.95 & \ldots & \ldots  \\
HIP 36758    &  07 33 34.77 &  $+$39 27 14.0 &  9.75 &    5.95 & F8    & V? &  11 &   1540 &    36.73 &   1.02 &    2.7  \\
HIP 38195    &  07 49 32.01 &  $+$41 28 08.3 &  9.41 &    8.98 & G5    & C? &  10 &   6508 &    71.24 &   1.69 &    7.1  \\
HIP 39681    &  08 06 34.41 &  $+$22 27 24.2 &  7.22 &   14.73 & G5IV  & S1 &  68 &  10984 &    $-$0.57 & \ldots & \ldots  \\
HIP 40253    &  08 13 17.31 &  $+$49 13 15.5 &  8.54 &    8.59 & F5    & S1 &  67 &   6961 &    $-$3.26 & \ldots & \ldots  \\
HIP 40724    &  08 18 44.42 &  $-$15 12 08.5 &  9.87 &   28.85 & K5V   & V  &   2 &    797 &    88.15 &   1.05 &    7.6  \\
HD 71028     &  08 26 07.19 &  $+$28 24 10.7 &  8.01 &    2.21 & K0III & S1 &  29 &   7645 &    34.15 & \ldots & \ldots  \\
HIP 42507    &  08 40 00.27 &  $-$06 28 33.1 &  9.88 &   38.80 & K6V   & C? &   5 &   2245 &   $-$12.14 &   1.38 &   17.1  \\
HIP 42550    &  08 40 22.54 &  $+$51 45 06.6 &  7.70 &    1.70 & G5III & C  &  11 &  10684 &  $-$101.16 &   0.44 &    1.3  \\
HD 73394B    &  08 40 18.25 &  $+$51 45 46.8 & 11.54 &    1.41 & --    & s2 &   5 &   4133 &   $-$32.54 & \ldots & \ldots  \\
HIP 43820    &  08 55 24.82 &  $+$70 47 39.2 &  8.61 &   86.32 & M1V   & V  &  19 &   5226 &    44.64 & \ldots & \ldots  \\
HD 75632B    &  08 55 24.82 &  $+$70 47 39.2 &  8.86 &   86.32 & M1V   & C  &   2 &    326 &    43.81 &   0.08 &    0.3  \\
HIP 46383    &  09 27 28.37 &  $+$39 30 17.9 &  9.85 &   30.60 & K4V   & S2 &  12 &   2585 &   $-$32.36 & \ldots & \ldots  \\
BD-08 2689   &  09 28 51.52 &  $-$09 16 00.8 & 10.54 &    9.20 & M0V   & V  &   4 &   2208 &   $-$14.81 &   1.55 &   24.5  \\
HIP 46926    &  09 33 52.46 &  $+$15 29 31.2 &  9.47 &    9.32 & G0    & S2 &  38 &   5046 &   $-$10.86 & \ldots & \ldots  \\
HIP 47133    &  09 36 15.91 &  $+$37 31 45.5 & 11.02 &   25.78 & M0V   & S2 &  16 &   5819 &    $-$0.68 & \ldots & \ldots  \\
HIP 47899    &  09 45 44.26 &  $+$50 14 08.3 & 11.19 &   12.58 & K4V   & S2 &  30 &   6588 &   $-$15.01 & \ldots & \ldots  \\
HIP 48346    &  09 51 18.98 &  $+$37 36 15.9 &  9.96 &   19.28 & K8    & S2 &  23 &   2828 &   $-$14.65 & \ldots & \ldots  \\
HIP 50156    &  10 14 19.18 &  $+$21 04 29.6 & 10.08 &   42.74 & M0.7V & S1 &  13 &   2527 &     8.74 & \ldots & \ldots  \\
HIP 50271    &  10 15 52.67 &  $+$25 50 01.2 &  9.44 &    5.77 & G0    & S1 &  49 &  10619 &     4.29 & \ldots & \ldots  \\
HIP 52021    &  10 37 47.41 &  $-$06 23 22.5 &  9.94 &   26.41 & K8    & V  &   8 &   6592 &   $-$11.58 &   7.11 &   99.0  \\
HIP 54002    &  11 02 50.12 &  $-$09 19 49.3 &  9.04 &   32.07 & K3V   & V  &   6 &   6554 &    $-$6.78 &   0.98 &    9.4  \\
HIP 54094    &  11 04 07.16 &  $+$53 22 55.5 &  9.98 &   21.45 & ?     & V? &   9 &   6554 &   $-$23.88 &   0.58 &    4.2  \\
HIP 56229    &  11 31 36.39 &  $+$40 30 01.2 &  9.75 &   22.85 & M0    & S2 &   8 &    473 &    12.02 & \ldots & \ldots  \\
HIP 57058    &  11 41 49.59 &  $+$05 08 26.5 &  9.59 &   32.26 & K4V   & S1 &  11 &   2602 &    18.90 & \ldots & \ldots  \\
BD+44 2120A  &  11 43 48.16 &  $+$44 10 46.5 & 11.22 &    2.86 & F5    & C  &   5 &   4136 &    $-$5.73 &   0.74 &    3.6  \\
BD+44 2120B  &  11 43 47.90 &  $+$44 10 40.3 & 11.45 &    3.06 & ?     & S1 &  17 &   4136 &    $-$4.39 & \ldots & \ldots  \\
BD+44 2120C  &  11 43 44.06 &  $+$44 10 49.1 & 11.04 &    3.06 & ?     & C  &  11 &   7654 &    15.01 &   0.57 &    0.6  \\
GQ Leo       &  11 47 45.73 &  $+$12 54 03.4 & 10.83 &   16.29 & K5Ve  & C  &  12 &   5468 &   $-$12.77 &   1.31 &   15.0  \\
HIP 57857    &  11 51 56.21 &  $+$33 07 11.4 & 10.97 &   18.50 & K0V   & V  &  11 &   6645 &   $-$20.28 &   2.96 &   95.6  \\
HIP 57949    &  11 53 05.24 &  $+$18 55 48.1 & 11.73 &   32.00 & M0.5V & V  &   7 &   5529 &     6.82 &   2.48 &   33.2  \\
HIP 59000    &  12 05 50.66 &  $-$18 52 30.9 & 10.02 &   42.72 & K5V   & -  &   1 &      0 &    $-$8.20 & \ldots & \ldots  \\
HIP 60433    &  12 23 26.85 &  $+$20 17 27.0 & 10.02 &   24.99 & K4V   & V  &  15 &   6594 &   $-$40.01 &   6.20 &   99.0  \\
HIP 60448    &  12 23 34.71 &  $+$27 54 47.6 & 11.41 &   33.82 & K5V   & V  &   5 &   6592 &   $-$31.14 &   1.39 &   22.8  \\
HIP 61436    &  12 35 19.72 &  $+$34 04 06.6 & 10.51 &   24.43 & K5V   & S2 &  24 &   2671 &    $-$0.01 & \ldots & \ldots  \\
HIP 62505    &  12 48 32.31 &  $-$15 43 10.1 &  7.89 &   39.21 & K2.5V & V  &  12 &   4417 &    $-$1.87 &   3.35 &   32.9  \\
HIP 62755    &  12 51 34.58 &  $+$59 50 25.9 & 11.31 &    0.81 & K?    & S1 &  11 &   3000 &   $-$47.03 & \ldots & \ldots  \\
HIP 63253    &  12 57 40.21 &  $+$35 13 30.1 & 10.68 &   46.84 & M0V   & V  &  11 &   2615 &   $-$10.00 &   1.32 &   25.7  \\
HIP 63816    &  13 04 46.60 &  $+$55 54 10.1 & 10.75 &   30.63 & M0V   & C  &   3 &   2598 &   $-$22.47 &   0.32 &    1.6  \\
HIP 63942    &  13 06 15.40 &  $+$20 43 45.3 &  9.40 &   53.18 & K4V   & C  &  13 &   5545 &    $-$2.84 &   0.57 &    1.3  \\
HIP 65012    &  13 19 34.69 &  $+$35 06 24.5 & 11.90 &   73.99 & M3V   & -  &   1 &      0 &    $-$5.20 & \ldots & \ldots  \\
HIP 65026    &  13 19 45.65 &  $+$47 46 41.0 &  8.76 &  109.98 & M2V   & S2 &  27 &   2629 &     0.12 & \ldots & \ldots  \\
HIP 65327    &  13 23 23.30 &  $+$57 54 22.1 &  9.56 &   41.47 & K5V   & S1 &  15 &   6115 &    $-$8.53 & \ldots & \ldots  \\
GJ 513       &  13 29 21.31 &  $+$11 26 26.5 & 11.92 &   52.30 & M3V   & V  &   4 &    416 &    29.55 &   2.73 &    8.5  \\
HIP 65887    &  13 30 22.60 &  $+$07 24 54.5 &  7.63 &    3.29 & K0    & S1 &  25 &   3674 &    $-$3.42 & \ldots & \ldots  \\
HIP 66290    &  13 35 11.42 &  $+$22 29 59.0 &  6.99 &   26.29 & F5V   & S1 &  36 &   8395 &   $-$14.04 & \ldots & \ldots  \\
HIP 67086    &  13 45 02.39 &  $+$02 05 31.5 & 10.78 &   21.67 & K5    & C? &  14 &   2886 &   $-$27.23 & \ldots & \ldots  \\
BD+26 2498   &  13 49 18.69 &  $+$25 52 54.2 &  9.85 &    1.40 & G5    & S1 &  31 &   6654 &   $-$13.64 & \ldots & \ldots  \\
HIP 67808    &  13 53 27.56 &  $+$12 56 33.4 &  9.77 &   45.62 & K7V   & V? &   6 &   2659 &   $-$18.71 &   0.43 &    2.2  \\
BD+19 2735   &  13 58 13.62 &  $+$19 17 11.8 &  9.31 &   27.17 & K2    & S1 &  20 &   5133 &    $-$9.53 & \ldots & \ldots  \\
HIP 68801    &  14 05 03.72 &  $+$10 00 48.9 &  8.68 &   19.58 & G5    & S2 &  24 &   5429 &   $-$36.32 & \ldots & \ldots  \\
HIP 69549    &  14 14 12.16 &  $+$18 05 06.8 &  7.98 &   11.72 & G     & S2 &  29 &   6951 &     8.58 & \ldots & \ldots  \\
HIP 71904    &  14 42 26.26 &  $+$19 30 12.7 & 10.03 &   42.24 & K5V   & C  &   2 &   2550 &   $-$28.08 &   0.09 &    0.2  \\
HIP 71914    &  14 42 33.81 &  $+$19 28 48.6 &  9.12 &   41.70 & K5V   & -  &   1 &      0 &   $-$28.70 & \ldots & \ldots  \\
HIP 72508    &  14 49 32.38 &  $+$51 22 28.2 &  6.48 &   19.09 & F5IV  & s2 &  11 &   7052 &    $-$6.13 & \ldots & \ldots  \\
BD+49 2364   &  15 13 25.00 &  $+$49 00 24.0 & 10.77 &    0.97 & ?     & V  &  22 &  10789 &   $-$74.33 &   1.69 &   19.7  \\
HIP 76941    &  15 42 38.57 &  $+$31 56 45.5 & 10.88 &   20.10 & K5V   & S1 &  26 &   2886 &   $-$12.50 & \ldots & \ldots  \\
HIP 77141    &  15 45 00.29 &  $+$35 57 40.9 & 10.11 &   18.50 & K4/5V & S1 &  30 &   2913 &    $-$3.46 & \ldots & \ldots  \\
HIP 78158    &  15 57 33.88 &  $+$34 32 22.8 & 10.83 &   19.30 & K5V   & S1 &  19 &   2748 &   $-$44.84 & \ldots & \ldots  \\
HIP 79796    &  16 17 05.39 &  $+$55 16 09.1 &  9.46 &   49.35 & M1.5V & S1 &  14 &   2895 &   $-$30.82 &  12.49 &   99.0  \\
HIP 80751    &  16 29 14.36 &  $+$23 46 33.9 & 10.08 &   30.84 & K5V   & S1 &  14 &   6554 &   $-$32.81 & \ldots & \ldots  \\
BD+52 1968A  &  16 39 13.76 &  $+$52 37 39.4 & 10.01 &   22.86 & K8    & V  &  13 &   2926 &    $-$8.55 &   3.30 &   99.0  \\
BD+52 1968B  &  16 39 13.10 &  $+$52 37 38.2 & 11.40 &   22.71 & --    & C  &   2 &      0 &    $-$8.69 &   0.39 &    6.9  \\
HIP 82506    &  16 51 46.46 &  $+$25 24 00.7 &  7.09 &   14.84 & F4III & V  &   8 &   4785 &   $-$25.39 &   4.76 &   99.0  \\
BD+61 1678C  &  17 35 34.49 &  $+$61 40 53.6 &  9.97 &   69.83 & M1V   & C  &   9 &   3105 &   $-$15.33 &   0.75 &    2.1  \\
HIP 86221    &  17 37 10.77 &  $+$27 53 47.2 &  9.20 &   32.00 & M0V   & C  &   2 &   2825 &   $-$42.55 &   0.33 &    3.3  \\
BD+27 2853C  &  17 37 11.41 &  $+$27 53 51.4 & 11.61 &   31.09 & K5    & -  &   1 &      0 &   $-$43.39 & \ldots & \ldots  \\
HIP 90274    &  18 25 10.11 &  $+$64 50 18.3 &  6.86 &    5.70 & K0    & s2 &  28 &   5881 &   $-$52.07 & \ldots & \ldots  \\
HIP 91043    &  18 34 20.10 &  $+$18 41 24.2 &  7.45 &   28.27 & G2V   & s2 &  39 &   4879 &   $-$22.21 & \ldots & \ldots  \\
HIP 92952    &  18 56 15.93 &  $+$54 31 48.1 & 10.37 &   21.51 & M0V   & S1 &  27 &   2944 &   $-$18.38 & \ldots & \ldots  \\
HIP 94557    &  19 14 39.16 &  $+$19 19 03.7 & 11.54 &   55.24 & M4.5V & V  &   3 &   2904 &    $-$4.28 &  19.34 &   99.0  \\
HIP 94622    &  19 15 18.84 &  $+$24 53 49.5 &  9.78 &   34.57 & M0    & s2 &   7 &   2989 &   $-$72.83 & \ldots & \ldots  \\
BD+77 767    &  20 11 00.50 &  $+$77 43 18.4 & 10.30 &   24.54 & K8    & S1 &  18 &   2822 &   $-$25.83 & \ldots & \ldots  \\
HIP 99969    &  20 16 55.43 &  $+$06 55 18.3 &  9.47 &   22.67 & K4V   & s2 &   8 &   2611 &   $-$56.38 & \ldots & \ldots  \\
HIP 101941   &  20 39 29.75 &  $+$28 05 18.6 &  7.78 &    2.63 & K4III & S1 &  46 &   4775 &   $-$20.56 & \ldots & \ldots  \\
HIP 102300   &  20 43 41.37 &  $+$64 16 54.1 & 11.38 &   46.53 & M0Ve  & V  &   5 &   2963 &    15.51 &   1.25 &   11.8  \\
HIP 102320   &  20 43 53.36 &  $+$31 19 10.4 &  9.96 &   24.06 & K5    & S2 &  29 &   3013 &   $-$21.08 & \ldots & \ldots  \\
HIP 102718   &  20 48 50.72 &  $+$05 11 58.8 &  9.69 &    9.70 & F7Vw  & C? &   6 &   4290 &  $-$117.67 &   0.64 &    2.8  \\
HIP 103375   &  20 56 37.80 &  $+$52 49 36.9 &  9.66 &    7.55 & G0    & C? &  11 &   4418 &   $-$47.55 &   0.53 &    2.5  \\
HIP 104994   &  21 15 54.95 &  $+$28 57 47.4 & 10.41 &    6.97 & G5    & S3 &  30 &   6669 &   $-$46.87 & \ldots & \ldots  \\
HIP 105504   &  21 22 07.78 &  $-$10 30 47.9 & 10.33 &   24.00 & K7    & V  &   5 &   2613 &     7.40 &  17.45 &   99.0  \\
BD+47 3439   &  21 30 48.00 &  $+$48 27 25.4 &  8.66 &    7.12 & K0    & s2 &  14 &   5465 &    10.04 & \ldots & \ldots  \\
HIP 110291   &  22 20 23.85 &  $+$46 25 05.7 &  8.51 &   16.12 & G0    & S1 &  40 &   1571 &  $-$108.04 & \ldots & \ldots  \\
HIP 110526   &  22 23 29.10 &  $+$32 27 33.9 & 10.76 &   64.47 & M3.0V & C  &  13 &   2940 &   $-$19.90 &   0.52 &    2.4  \\
HIP 110978   &  22 29 04.21 &  $-$13 42 01.8 &  9.17 &    1.25 & K2III & V? &   3 &     34 &    $-$1.37 &   0.82 &    6.3  \\
HIP 111685   &  22 37 29.90 &  $+$39 22 51.6 &  9.41 &   41.99 & M0Ve  & S1 &  10 &   5457 &   $-$58.38 & \ldots & \ldots  \\
HIP 111942   &  22 40 30.30 &  $+$43 00 47.4 &  9.83 &   31.92 & K8V   & S1 &  20 &   2940 &   $-$31.76 & \ldots & \ldots  \\
HIP 112040   &  22 41 34.99 &  $+$18 49 27.5 & 10.51 &   31.85 & M0V   & V  &  13 &   5469 &   $-$15.94 &   1.17 &   10.9  \\
HIP 112268   &  22 44 24.92 &  $+$17 33 23.8 & 10.19 &   19.83 & K6V   & V  &  13 &   5521 &   $-$38.11 &   0.95 &    8.7  \\
HIP 112523   &  22 47 30.45 &  $+$19 13 27.4 & 10.29 &   26.95 & K5V   & S1 &  13 &   5458 &     3.97 &   1.86 &   28.7  \\
HIP 116003   &  23 30 13.44 &  $-$20 23 27.5 & 11.11 &   62.67 & M2Ve  & V  &   2 &      7 &    $-$6.83 &   7.07 &   91.2  \\
BD+66 1664   &  23 58 10.67 &  $+$67 33 59.7 &  8.73 &   11.13 & G5    & S1 &  33 &   1175 &   $-$22.08 & \ldots & \ldots  \\
HIP 118212   &  23 58 43.51 &  $+$46 43 44.9 &  9.62 &   58.41 & K7V   & S1 &  18 &   2986 &     1.00 & \ldots & \ldots  \\
\end{longtable}
}

The  core of  the observing  program is  the survey  of K-  and M-type
dwarfs  featured  in  the  McCormic  catalog \citep[see  a  review  in
][]{MCC}  and in the  catalog of  nearby stars  \citep{Gliese}.  Stars
with constant  RVs, presented by  \citet{S16}, were used to  study the
local  kinematics.   Objects with  variable  RVs,  studied here,  were
monitored more extensively for orbit determination.  The spectroscopic
survey  of  nearby  K  dwarfs  by  \citet{Halbwachs2003,Halbwachs2018}
pursued similar  goals, and six   of their objects  are common to
our  sample.   In addition  to  the nearby  dwarfs, we  monitored
  several  other objects  with  variable RVs  and  present here  their
  orbits.   In particular,  our program  was  augmented by  the {\it
  Hipparcos} stars  with astrometric accelerations  \citep{MK05}, with
the aim to establish their periods. 

Table~\ref{tab:list} contains the object  list and the synopsis of our
results.  Its  first column gives common identifiers  (HIP numbers are
preferred, with HD or BD as  the second choice), and the following two
columns  give the equatorial  coordinates for  J2000. Then  follow the
visual magnitude $V$, parallax $\varpi$, and spectral type.
All  these  data  are  recovered  from Simbad.   Most,  but  not  all,
parallaxes come  from the {\it Gaia} DR2  \citep{Gaia}.  The remaining
columns of Table~\ref{tab:list} summarize our results. The variability
is coded as C -- constant,  V -- variable, s2 -- double-lined, S1, S2,
or  S3  --  single-  double-  and triple-lined  binaries  with  orbits
determined here. Then follow the  number of RV measurements $N$, their
time span $\Delta  T$, and the weighted mean  velocity $\langle RV \rangle$.
For binaries  with orbits, $\langle RV \rangle$  is the center-of-mass
velocity.   The  last two  columns  contain  the statistics  explained
below.

  \begin{figure}
   \centering
   \includegraphics[width=\hsize]{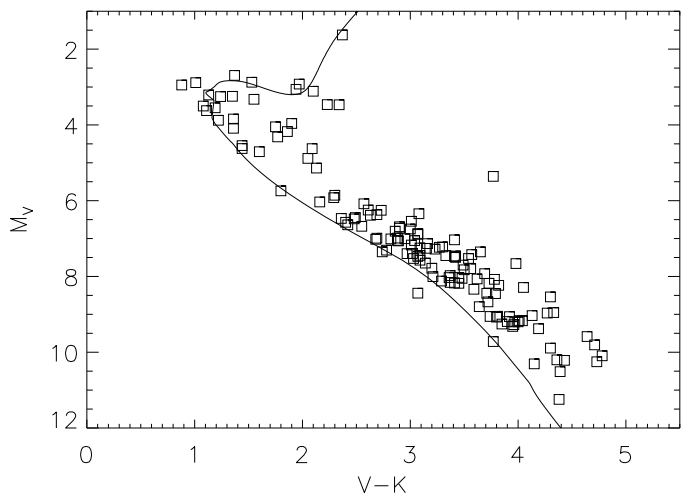}
      \caption{Color-magnitude      diagram     of      stars     from
        Table~\ref{tab:list}. The line is  a 4-Gyr isochrone for solar
        metallicity from  \citet{PARSEC}. The discrepant  point is
        caused by the erroneous  parallax of BD$-$08~2689.  }
         \label{fig:cmd}
   \end{figure}

To  give an  idea of  the stars  studied here,  we place  them  on the
color-magnitude diagram in Fig.~\ref{fig:cmd}. Most stars are low-mass
dwarfs, although  hotter F-type stars  and giants are also  present in
our  sample.   Remember  that  at   the  start  of  our  program,  the
trigonometric parallaxes were not available, and some stars classified
spectroscopically  as dwarfs  turned  out to  be  giants.  The  median
parallax is 25\,mas, so half  of the objects are located within 40\,pc
from the  Sun. The closest, HIP~29295,  has a distance of  only 5.7 pc.
On the  other hand,  16 objects (mostly  giants) have  parallaxes less
than 5\,mas.

\section{Observations and data processing}
\label{sec:inst}

\subsection{Instruments}

The  first  RV  measurements  reported  here  date  back  to  February
1988. They were made using the correlation radial-velocity meter (RVM)
installed  at  the 1-m  Lithuanian  telescope  at  Mt.  Maidanak,  in
Uzbekistan  \citep[see e.g.][]{Tok1992}.   Like the  CORAVEL instrument
\citep{CORAVEL}, it scans the  echelle spectrum over the physical mask
with slits corresponding to individual spectral lines, accumulates the
transmitted flux as  a function of the relative  shift, and determines
the  RV  by approximating  the  cross-correlation  curve  with one  or
several Gaussian curves.

Starting from 1998,  a similar CORAVEL-type instrument
constructed at the Vilnius  observatory became operational. It worked,
mostly, at  the 1.65-m telescope of the  Moletai observatory, although
several  trips to  other  telescopes were  made.   The instrument  and
observing runs are further  described by \citet{S16}.  That paper also
gives  a thorough  analysis  of the  RV  zero points  and accuracy  by
comparing to several lists of RV standards. In the following, we refer
to both  instruments as  CORAVELs, without making  distinction between
them.

 \begin{figure}
   \centering
   \includegraphics[width=\hsize]{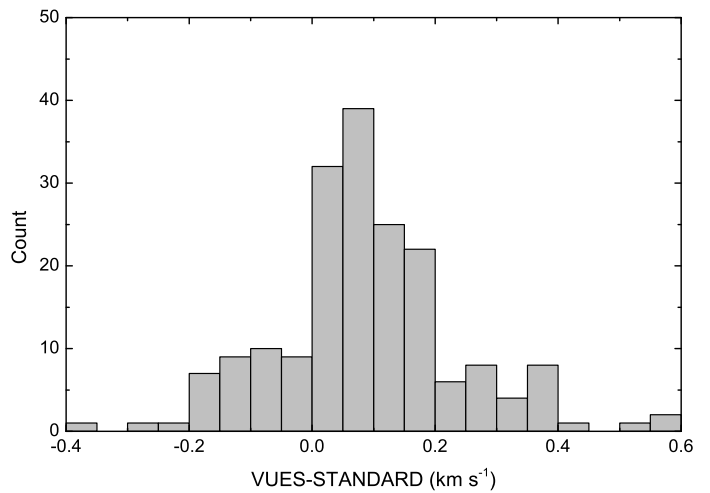}
      \caption{Radial velocities of IAU  RV standard stars measures by
        VUES. The histogram of the RV difference is plotted. }
         \label{fig:stand}
   \end{figure}

In 2015, the  CORAVEL in Moletai was replaced  by the modern fiber-fed
echelle spectrometer VUES \citep{VUES}.   It covers the spectral range
from 400\,nm  to 880\,nm  with a resolutions  from 30000 to  60000. In
this program  we used  the lowest resolution  of 30000.  The  first RV
measurement with VUES reported here  was made on November 26, 2015 (JD
2457352).  The spectrum recorded by  the CCD detector is extracted and
calibrated in  the standard  way.  The RV  is determined  by numerical
cross-correlation of  this spectrum with a binary  mask, emulating the
CORAVEL method in software. Compared  to CORAVEL, the RVs delivered by
VUES  are more  accurate; their  rms  residuals from  the orbits  are,
typically, from 0.2  to 0.3 \kms.  The RV zero  point is controlled by
observations  of  the IAU  RV  standards (Fig.~\ref{fig:stand}).   The
systematic RV offset of VUES is less than 0.1 \kms.

\subsection{Detection of variable RVs}

For  each star,  the  mean  radial velocity  $\langle  RV \rangle$  was
computed  with weights inversely  proportional to  the squares  of the
measurement errors. The errors of the CORAVEL RVs $\sigma_i$ were used
as listed, while the errors of the RVs measured by VUES were augmented
by  adding  quadratically 0.2  \kms  because  the  listed errors  are
internal, determined  by the dip fitting,  and they do  not account for
other  error sources  such  as wavelength  calibration and  instrument
stability.

The weighted rms deviation from the mean, $\sigma_V$, is computed as
\begin{equation}
\sigma_v = \sum_{i=1}^N (RV_i - \langle  RV \rangle)^2 \sigma_i^{-2}
\; / \;  \Sigma_{i=1}^N \sigma_i^{-2}  .
\label{eq:sigma}
\end{equation}
The first term of this equation, divided by the number of measurements
$N-1$,  gives the  normalized  $\chi^2/(N-1)$ statistic,  which has  a
mathematical expectation of one for  a constant RV and realistic errors
$\sigma_i$. The statistics $\sigma_V$  and $\chi^2/(N-1)$ are given in
the  last two columns  of Table~\ref{tab:list},  except for  the stars
with  computed  orbits.  The  large  values  $\chi^2/(N-1)  >100$  are
replaced by  99.  For multi-lined systems, the  statistics are computed
for the primary component.

\subsection{Orbit calculation}

Spectroscopic orbits  were determined  with the help  of the  IDL code
{\tt  orbit.pro}.\footnote{The  code  can  be  downloaded  from  \url{
    http://www.ctio.noao.edu/~atokovin/orbit/}}   Individual  RVs  are
weighted in  proportion to  $\sigma_i^{-2}$.  However, the  errors are
artificially increased when RVs are deduced from partially blended dips and
in other instances were the residuals strongly exceed the errors. In a
few cases  where the  spectroscopic pair is  also resolved,  we fitted
combined spectro-visual orbits using the  same code. The errors of RVs
and  positional  measurements  are  balanced  in the  sense  that  the
normalized statistic $\chi^2/N$ for each  type of data should be close
to  one.  Finally,  orbits  of  triple systems  were  fitted using  an
extension  of  this code  called  {\tt  orbit3.pro}  and described  by
\citet{TL2017}; it is also available online.

\section{Results}
\label{sec:res}

\subsection{Individual RVs}

\begin{table}
\caption{Radial velocities (fragment) \label{tab:rv}  }
\centering
\begin{tabular}{l c c c c c}
\hline\hline
Name & JD  & RV  & $\sigma$ & Inst. & a/b \\
     & +2400000 & (\kms) & (\kms) &  & \\
\hline
HIP 96     &    55470.486 & $-$11.70 &   0.60 &C &\\
HIP 96     &    55485.410 & $-$11.60 &   0.50 &C &\\
HIP 3428    &   58387.512 & $-$1.09  &   0.42 & V & a \\
HIP 3428    &   58387.512 & $-$14.29 &   0.53 & V &  b \\
\hline
\end{tabular}
\end{table}

Table~\ref{tab:rv}, published in full  at the CDS, lists individual RV measurements, a total of 1913 velocities
obtained with CORAVEL and 632 velocities measured with VUES. Its first column is the object  name (same as  in the
object list). Then follow the Julian date, RV, its internal error, and the instrument code  (V for VUES and C  for
CORAVEL). For double-lined binaries,  the last column distinguishes the primary and  secondary components by  the
letters a and b, respectively.  For  VUES, we list the internal errors determined  by fitting the correlation
dips, while for CORAVEL the errors include the instrumental noise.

\subsection{Spectroscopic orbits}

The   orbital    elements   and    their   errors   are    listed   in
Table~\ref{tab:sborb}, in  standard notation. For  circular orbits, we
fixed   the  eccentricity   $e$   and  the   argument  of   periastron
$\omega$. The before-last column  gives the weighted rms residuals for
the  primary component  or for  both components  of  double-lined (SB2)
binaries.  The  spectroscopic  masses  $M_{1,2} \sin^3  i$,  i.e.  the
minimum  masses, are provided  for SB2s  in the  last column.  For the
single-lined pairs (SB1s), this  column contains the minimum secondary
mass estimated  from the orbit  after adopting a reasonable  guess for
the primary mass.

The   RV    curves   of   spectroscopic   binaries    are   given   in
Figs.~\ref{fig:orb1}--\ref{fig:orb3}.  In  each panel, the horizontal  axis is the
orbital phase  from 0 to 1.5  (the last half-period  is repeated), the
vertical axis  is the RV  in \kms. The  RV curves of the  primary and
secondary   components  are   plotted  in   full  and   dashed  lines,
respectively, while the squares  and triangles denote the measurements. In
some plots, crosses denote RVs with reduced weights.

 {  \begin{figure*}
   \centering
  \includegraphics[width=\hsize]{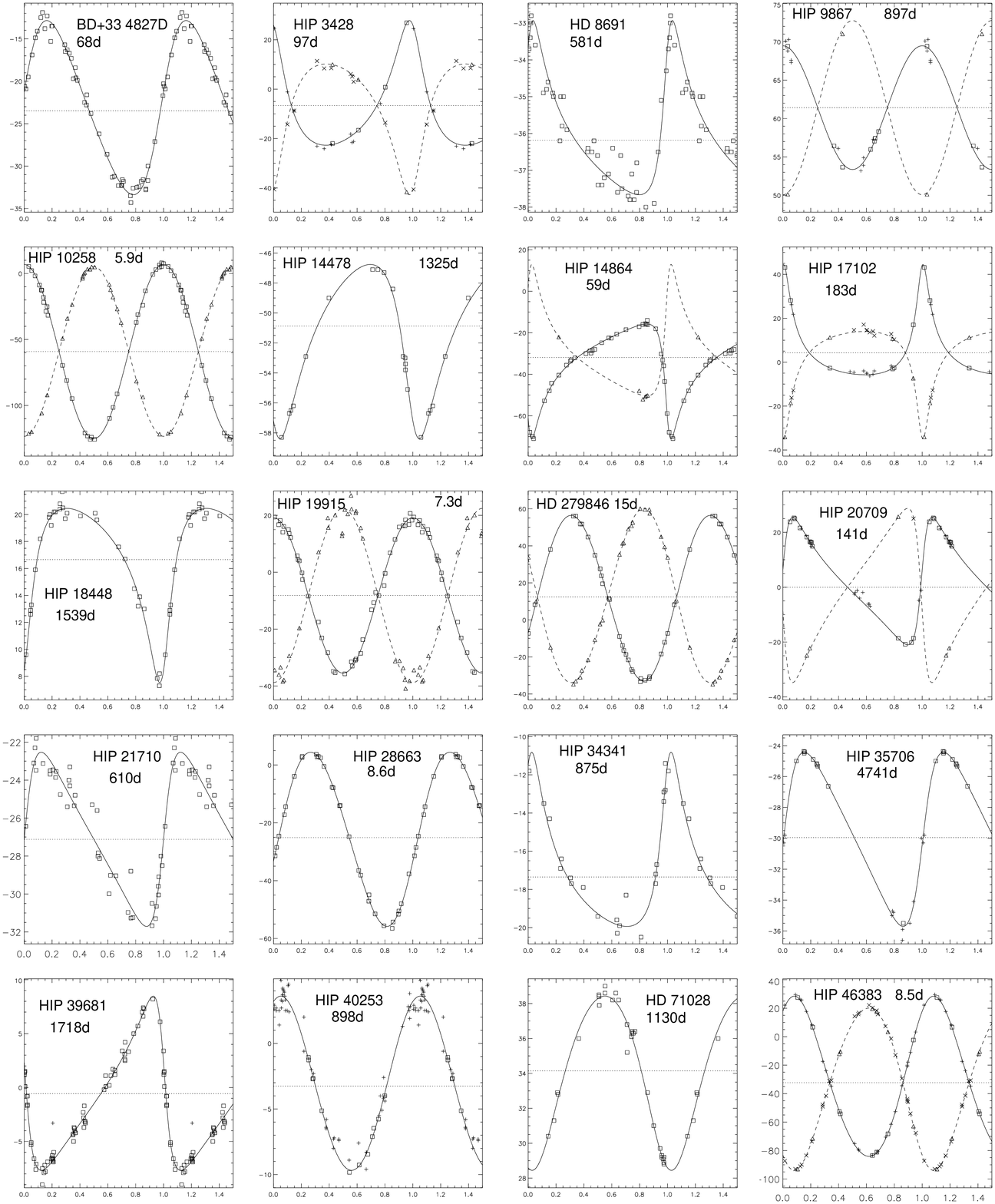}
   \caption{RV  curves.  Object  names  and  approximate  periods  are
      indicated.  In each  plot,  the horizontal  axis  is the  orbital
     phase, the vertical axis is the  RV in \kms. The RV curves of the
     primary and  secondary components are plotted in  full and dashed
     lines,  respectively,  while  the  squares and  triangles  denote
     the measurements.  In some  plots,  crosses denote  RVs with  reduced
     weights. }
   \label{fig:orb1}
    \end{figure*}

   \begin{figure*}
   \centering
  \includegraphics[width=\hsize]{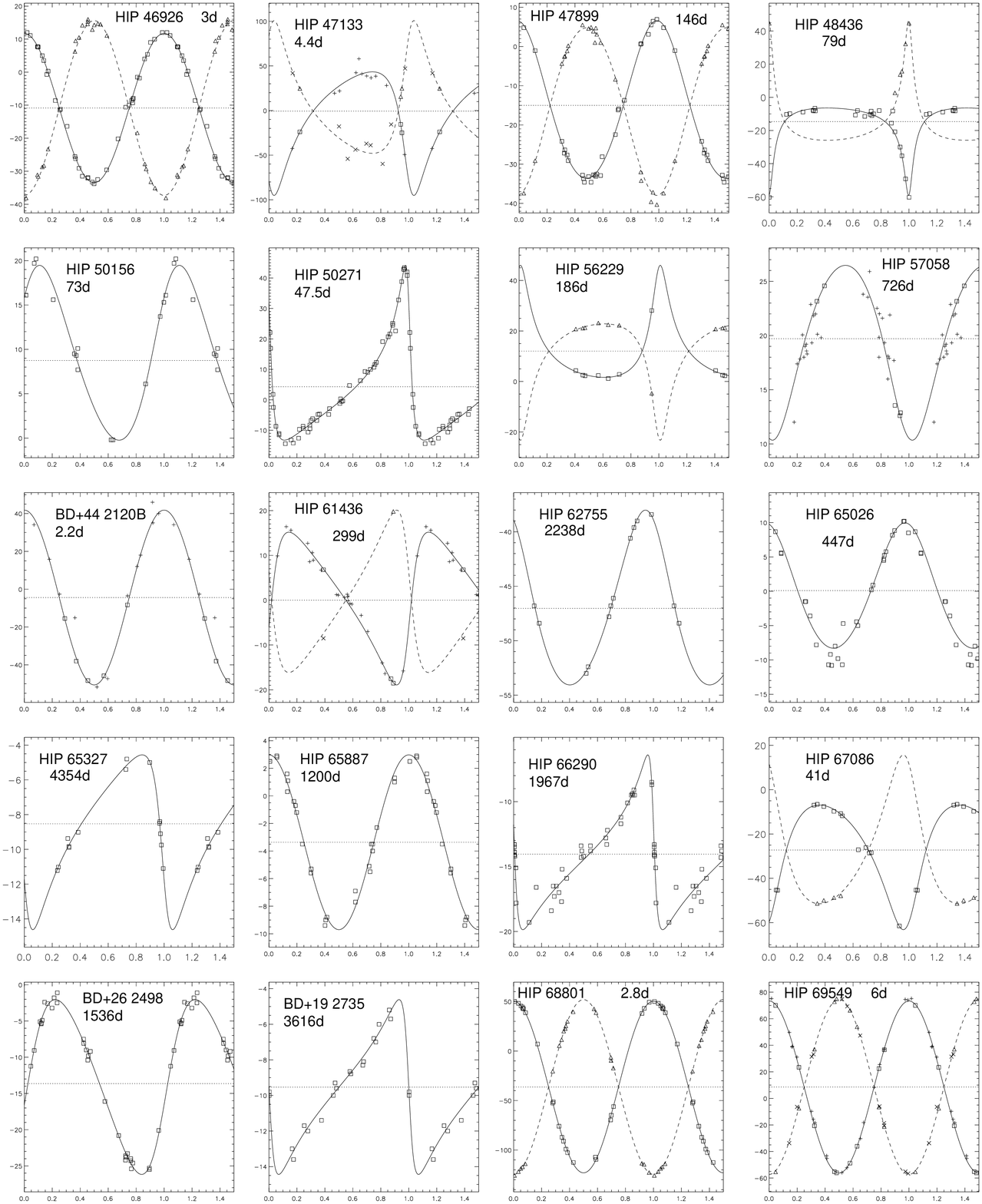}
   \caption{RV curves (continued). }
   \label{fig:orb2}
    \end{figure*}

   \begin{figure*}
   \centering
  \includegraphics[width=\hsize]{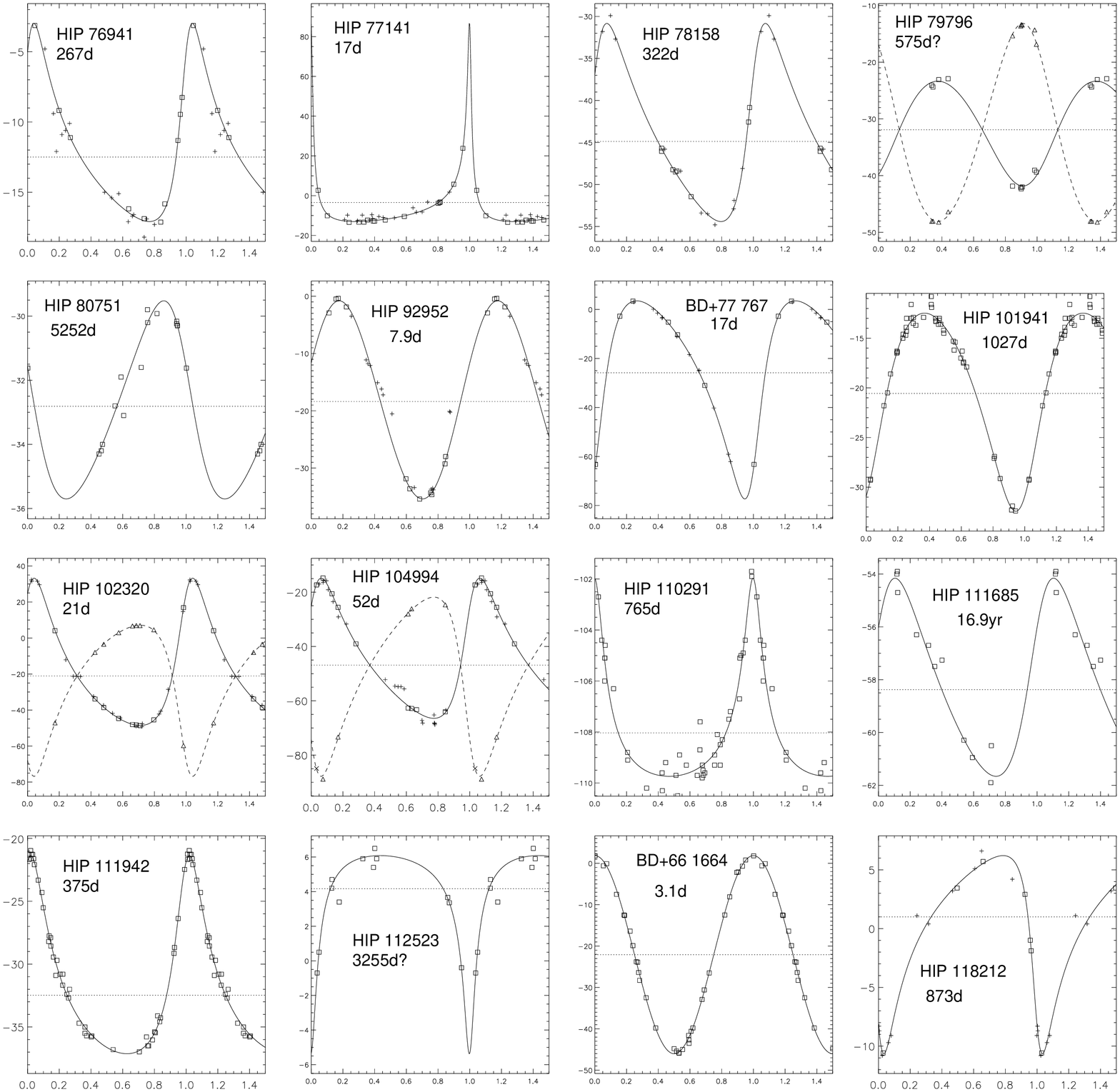}
   \caption{RV curves (continued). }
   \label{fig:orb3}
    \end{figure*}

\subsection{Comments on the individual objects}

This subsection provides notes on individual stars and stellar systems
from our list.  Hierarchical systems with three or more components are
also featured  in the Multiple-Star Catalog,  MSC \citep{MSC}. Several
objects with subsystems discovered here are added to the MSC.  Data on
visual components  are taken from the Washington  Double Star catalog,
WDS \citep{WDS}  and from the MSC.\footnote{See the latest version
    at http://www.ctio.noao.edu/\~{}atokovin/stars.} Unknown periods
of visual  binaries are  estimated crudely from  projected separations
assuming  that they  equal  the semimajor  axis.    Similary,  the
  semimajor axes  of spectroscopic  binaries are estimated  from their
  periods,  using  known  distance  and  a guess  of  the  components'
  masses.   Information on  astrometric accelerations detected by the
{\it Hipparcos} mission comes from the paper by \citet{MK05}.

{\bf HIP  96} (BD+13~5195, M0.5V,  43 pc).  This visual  triple system
 consists of  the 11\arcsec ~pair A,B and  the 0\farcs2 subsystem
Aa,Ab with $P \sim 25$ yr.  All components are M-type dwarfs.  We find
that the RVs of both A and B are slowly variable. This is expected for
A, which  is a close  pair; the component  B may also host  a low-mass
companion.

{\bf HIP 374} (HD 225220), a K0  giant, is the main component A of the
hierarchical system located  at the 200 pc distance.  The outer
pair  A,D has  a 95\farcs3  separation; A,B  is a  visual  binary with
$P=545$ yr. The main-sequence star D (TYC~2267-1300-1, probably of F9V
spectral type)  is found  here to be  an SB1  with $P=68$ d.   This is
therefore a quadruple system of  2+2 hierarchy.  The star C = HIP~375,
listed in the WDS, is optical, as evidenced by its proper motion (PM),
different parallax, and the RV measured here.

{\bf HIP 1412} (K7V, 32 pc) has a large RV variation, but no orbit can
be derived yet from our 6 RVs.

{\bf HIP  3428} (BD+23~97,  K7, 45 pc)  is a double-lined  twin binary
with $P=97$ d and the mass ratio $q = 0.96$.  Interestingly, it has an
astrometric acceleration, so, likely,  it is a triple system. However,
residuals to the spectroscopic orbit do not show any slow trends.

{\bf HIP 5110 } (HD 6440,  27 pc) is a 6\farcs1 physical binary STF~87
with  an  estimated period  of  $\sim$2  kyr.  We  discover  RV
variability  of the component  B, of  K8V spectral  type.  Considering
also the 2.5 \kms RV difference  between A and B, presumably caused by
the orbital motion of Ba,Bb, we believe that this is a triple system.

{\bf HD 8691} (G0, 50 pc) is a high-PM star and an SB1 with $P=581$ d,
with a low-mass secondary.

{\bf  HIP  9867}  (GJ~84.2, M0V,  19  pc)  is  a  high-PM star  and  a
double-lined  pair with $P=897$  d and  unequal correlation  dips. WDS
lists three visual companions, all  optical. Eclipses are
reported by \citet{Malkov2006}.

{\bf HIP 10258 }  (BD+03~301,  K5,  47  pc)  is a  chromospherically  active
double-lined binary with $P=5.9$ d and a mass ratio $q=0.93$. Its
visual companion at 25\arcsec (SKF~1518) shares common parallax, PM,
and RV.

{\bf  BD+49  646}  (unknown  spectral   type,  53  pc)  has    double
correlation  dips,  but  we have  not  derived  its  orbit from  the  8
spectra. It is an X-ray source.

{\bf  HIP 11437} (AG Tri, K7V,  41 pc)  is a  young chromospherically
active star in  the $\beta$~Pictoris moving group \citep{Messina2017};
it has extensive  coverage in the literature.  We  found a constant RV
of 6.0 \kms \citep[see also][]{S16}, in agreement with other published
studies.  The visual companion at 22\arcsec ~is physical.

{\bf HIP 12787  } (MCC 401, M0Ve,  49 pc) is a triple  system with the
outer  21\arcsec physical  binary A,C.   Its primary  component  is an
astrometric  binary,  resolved  directly  in 2015.9  at  0\farcs25  by
\citet{Janson2017}; its estimated period is $\sim$40 yr.  The spectrum
is  double-lined, suggesting existence  of a  close subsystem,  but no
orbit is derived.   The components A and C are  located above the main
sequence and belong to the  $\beta$ Pictoris moving group according to
Janson  et al.  The component  B at  25\arcsec ~listed  in the  WDS is
optical.

{\bf HIP 13398} (G~36-38, M2V, 23 pc) is a high-PM star, likely with
a variable RV.

{\bf HIP 13460 } (BD+60~585, K3V, 39  pc) is a triple system discussed
in the next subsection.

{\bf HIP  14478} (V568~Per, K6, 27  pc) is a triple  system. The outer
2\farcs9  pair A~1572  has an  estimated period  of $\sim$600  yr. Its
primary  component is a  single-lined binary  with $P=1325$  days. The
semimajor  axis  of the  inner  subsystem  is  87\,mas and it is
detectable astrometrically from the PM difference between {\it Gaia}
and {\it Hipparcos}.

{\bf HIP  14669 }  (MCC~99, M2V,  17 pc) is  a {\it  Hipparcos} visual
binary with  known orbit, $P=28.3$ yr.  We  see an RV trend  by 6 \kms
over 7 years, presumably caused by this orbit.

{\bf HIP 14864} (BD+60~637, M0Ve, 25 pc) is a triple system consisting
of the 0\farcs6 binary  discovered by {\it Hipparcos} (period $\sim$40
yr) and the double-lined subsystem Aa,Ab with $P=59.5$ d and $q=0.88$,
discovered here.

{\bf  BD+03~480} (V1221  Tau,  G0, 83  pc)  is a  young visual  triple
system, where the inner 0\farcs9  binary A~2417BC has been known for a
long  time (since  1912), while  another companion  D at  2\farcs2 was
discovered a century later, in 2012, and has not yet been confirmed as
physical (it  is not  found in {\it  Gaia} DR2).  Our  11 observations
during 13 years show a constant  RV of 12.66 \kms. However, {\it Gaia}
DR2 gives an  RV of 18.05 \kms with an error  of 7.26 \kms, suggesting
variability.

{\bf HIP 17102} (HD 278874, 39 pc)  is a flaring K2V dwarf in a triple
system. The  outer pair ES~327 has  a 15\farcs5 separation  and a long
$\sim$10 kyr period.   The secondary star B is  located above the main
sequence.  The main component A  is a double-lined binary with $P=183$
d  and $q=0.96$  (the components  Aa and  Ab are  interchanged  in our
orbit).  The  inner semimajor axis is  18 mas, so  the subsystem Aa,Ab
can be resolved.

{\bf GJ 3248}  is an M1V dwarf at 16\,pc. We  suspect that its RV is variable.

{\bf HIP 18448} (K0, 149 pc)  is a triple system composed of the outer
25\arcsec  pair LDS~1583  and the  inner subsystem  Aa,Ab  revealed by
astrometric acceleration.   Here we derive  its accurate spectroscopic
orbit with  $P=4.2$ yr. The primary  star is a  subgiant located above
the main sequence, while the component B is below; \citet{Chaname2012}
consider B to be a white dwarf.

{\bf HIP 19140} (BD$-$15~728, K5V, 40 pc) certainly has a variable RV,
as well as astrometric acceleration.

{\bf HIP 19915} (HD 26872, F8, 166 pc) is a triple system composed of
the tight  32-mas interferometric pair YSC~128 with  an estimated period
of $\sim$6 yr  (no orbit is known yet)  and the double-lined subsystem
with $P=7.3$  days. The inner orbit  is seen at  large inclination, as
evidenced by the small spectroscopic masses $M \sin^3 i$. We could not
detect variations of the systemic velocity that might be caused by the
visual binary.

{\bf  HD  279846} (K2,  82  pc) is  just  a  double-lined binary  with
$P=15.5$ days and an accurately determined orbit; $q=0.95$.

{\bf HIP 20709} (HD~27961, F5,  132 pc) is an interesting hierarchical
system where both  the outer 82-yr visual orbit  and the inner 141-day
double-lined  orbit,  determined here,  are  known.   The spectrum  is
triple-lined.   We fitted both  orbits simultaneously,  accounting for
the slow  RV drift  caused by the  visual pair HU~609.   The inclination 
of the spectrosopic pair derived from comparison between the spectroscopic mass $M \sin^3 i$ and 
the mass estimated from absolute magnitude, is  60\degr  or 120\degr, 
 while the outer inclination
is 122\degr.  The two orbits thus can be coplanar. The inner semimajor
axis  is 5\,mas,  so the  system  can be  resolved with  long-baseline
interferometers like CHARA array  to measure the relative inclination.
The rms scatter of RVs of the visual secondary component, B, is large,
2.3  \kms.  A possible  orbit of  Ba,Bb with  a 1600  d period  and an
amplitude of 2.3 \kms can be  fitted, reducing the weighted rms to 0.6
\kms. This tentative orbit is not given here.  The minimum mass of the
hypothetical component Bb is 0.15 \msun.

{\bf HIP 21710 } (HD 286955,  K2, 27\,pc) is a nearby 3-tier quadruple
system. The outer 34\arcsec pair  A,B (GIC~51) has an estimated period
of  23 kyr,  the intermediate  visual  binary Aa,Ab  resolved by  {\it
  Hipparcos} has an orbit with $P=204$ yr, and the inner spectroscopic
binary has a period of 610 days announced by \citet{Halbwachs2003} and
eventually  published  by  \citet{Halbwachs2018}.  We  determined  the
spectroscopic orbit  from our own observations, but  publish here more
accurate elements derived from the combined data.

{\bf  HIP  21845 }  (HD  29696,  F8, 115  pc)  might  have  a slow  RV
variation, although comparison between  {\it Gaia} and {\it Hipparcos}
does not reveal any astrometric acceleration.  The visual companion at
29\arcsec ~listed in the WDS  is physical, according to its {\it Gaia}
parallax and the RV of 11.4 \kms (the first discordant measure of this
pair given in the WDS is misleading).

{\bf HIP 23550} (HD 32387, G8V, 73 pc) shows an RV trend over 9 years,
indicative of  a long  period; it is  an acceleration binary.   The RV
variability  with   an  rms   of  1.8  \kms   was  also   detected  by
\citet{Niedever2002}.

{\bf HIP 24488 } (HD 33798, V390 Aur) is a visual binary with a period
of 513 yr according to  its current, still uncertain, orbit.  The {\it
  Gaia}  parallax of  2.14 mas  is erroneous,  so the  {\it Hipparcos}
parallax  of  8.73  mas  is  adopted, in  better  agreement  with  the
dynamical parallax from the orbit, 6.0 mas.  \citet{Fekel1991} studied
this  lithium-rich chromospherically  active G5III  giant  and concluded
that it is not a spectroscopic  binary; they quote the RV of $22.5 \pm
0.2$ \kms.  We  noted doubling of some correlation  dips that could be
caused by  the fast rotation,  $V \sin i  = 29$ \kms. The  RVs derived
from double dips are ignored here, and the RV is likely constant.

{\bf GJ 220} is a nearby  M2V dwarf (parallax 51.5$\pm$4.6 mas) with a
variable RV. We derive a tentative  orbit with $P=700$d and $K_1 = 2$
\kms from the 13 measured RVs.  More observations are needed, however,
to constrain the orbit before it can be published. The pair may have
been  resolved by  {\it Gaia}  because the  DR2 does  not  provide its
parallax.

{\bf HIP  28663} (HD~41028, F4IV, 103  pc) is a  single-lined binary
with a well-determined  orbit of $P=8.55$ d and  the minimum secondary
mass of 0.35 \msun.

{\bf HIP 29295}  (HD 42581, GJ 229) is a flaring  M1V dwarf located at
5.8 pc distance.  Its RV is  constant during 15.8 years spanned by our
observations.  Extensive literature  covers the search for exo-planets
with precise RVs and photometry  and the distant brown dwarf companion
GJ~229B detected by imaging.

{\bf HIP  29316} (BD+10~1032)  is an M3V  nearby (11 pc)  close visual
binary KAM~1 with  an estimated period of $\sim$20  yr. {\it Gaia} DR2
does  not provide  astrometry  of  this resolved  source.   The RV  is
variable, possibly because  of the visual orbit.  The  WDS companion C
at 13\arcsec ~is optical.

{\bf HIP 30269 }  (HD 44517,  F5V, 311  pc) has  a  large-amplitude RV
variation discovered by \citet{N04}, but no orbit yet.

{\bf HIP  33560 } (HD  51849, K4V, 22.5pc)  is a visual  triple system
composed of the 50\arcsec outer  pair AB,C and the 0\farcs6 inner pair
A,B discovered by {\it Hipparcos} with an estimated period of $\sim$50
yr  but yet unknown  orbit. We  suspect RV  variability that  might be
caused by motion in the visual  pair. The RV trend is also detected by
\citet{Halbwachs2018}.  The star is featured in \citet{S16}.

{\bf HIP  34341 } (BD+03~1552, K5V, 26  pc) is a  single-lined binary
with $P=875$ d, as well as an astrometric binary.

{\bf HIP  35706 }  (BD+68~474,  K5V, 42 pc)  has a  long period  of 13
years, fully covered by our RV data. The minimum mass of the secondary
is  rather large, 0.5  \msun, and  its lines  are likely  blended with
those  of  the primary,  reducing  the RV  amplitude.  It  is also  an
astrometric binary.

{\bf HIP 36758 } (BD+39 1967, F8, 168  pc) might have a variable RV, although
our 11 measurements are not conclusive.

{\bf  HIP 38195}  (HD 63207,  G~111-38, G5,  111 pc)  is  a three-tier
visual quadruple system with  separations of 109\arcsec, 2\farcs2, and
0\farcs084.  It  is metal-poor.  One of  our 10 RVs  deviates from the
rest,  suggesting variability.   However,  \citet{Latham2002} found  a
constant  RV  of 71.61  \kms,  so  the  existence of  a  spectroscopic
subsystems is  unlikely.  The  inner pair has  an estimated  period of
$\sim$20 years and should cause slow RV changes.

{\bf HIP 39681} (HD 66948, G5IV,  68 pc) is a single-lined binary with
a long 4.7-yr period. Astrometric orbit with similar period has been
published by \citet{Goldin2007}. The large minimum secondary mass, 0.5
\msun, indicates that the RV amplitude could be reduced by line blending.

{\bf HIP 40253 } (HD 68119, F5, 116 pc) is a single-lined  binary with
$P=808$ d, as well as an astrometric binary.

{\bf  HIP  40724  }  (BD$-$14~2469,  K5V, 35  pc)  has  only  two  RV
measurements that differ by 2 \kms; it is an astrometric binary.

{\bf HD 71028}  is a distant  (480 pc) chromospherically  active K0III  giant
 for which we determine an orbit with $P=1130$ d.

{\bf HIP 42507 } (BD$-$05~2603, K6V, 26 pc) has a large $\chi^2/(N-1)$
caused by  one outlying measurement,  hence the RV  variability  is not
certain.

{\bf HIP  42550} (HD~73394) is  a distant (600  pc) G5III  giant. Its
visual companion  B (ES~209) at 57\arcsec ~separation  is optical, with
different PM and RV. Both stars  were observed here, and we found that
B  has double lines.  No orbit  can be  derived  from  our five
spectra.

{\bf  HIP 43820}  (HD~75632, M1V,  11.6 pc)  is a  visual binary  on a
609-yr  orbit,  currently at  3\arcsec  ~separation.   The  RV of  the
brighter component A varies slowly  during 14 years of our monitoring,
while  its  mean  value differs  slightly  from  the  two RVs  of  the
component  B.   Therefore  we  believe  that  it  could  be  a  triple
system. \citet{Halbwachs2018} published 9  pairs of RVs, splitting the
CORAVEL correlation dips in two components.  Combining these data with
our RVs and  with two RVs from \citet{TS02}, we can  fit an orbit with
$P=1050$ d,  $K_1 = 2.8$ \kms, $K_2  = 7.0$ \kms, and  $\gamma = 44.6$
\kms. The minimum masses are  suspiciously small, 0.05 and 0.02 \msun,
therefore we prefer not to publish this orbit.

{\bf HIP 46383}  (BD+40~2208, K4V, 32 pc) is  a double-lined pair with
$P=8.5$\,d  and  equal  components,  $q=0.99$.  After  the  orbit  was
computed from our  12 RVs, the paper by  \citet{Halbwachs2018} came to
our  attention. Here  we  fit the  orbit  to the  combined  set of  33
RVs. The  small residuals of  0.38 and 0.43  \kms for the  primary and
secondary components, respectively, attest the good quality of our RVs
and  the  lack  of  substantial  zero-point  differences  between  the
CORAVELs at Moletai and OHP and the VUES. Considering the physical
companion at 56\arcsec (LEP 36), this is a triple system.

{\bf BD$-$08 2689} (M0V) shows an  increasing RV during 6 years of our
monitoring. Simbad quotes only  a crude parallax of 9.2$\pm$15.0 mas
that  places  the  star  at   3  mag  above  the  main  sequence  (see
Fig.~\ref{fig:cmd}). Most likely this parallax is wrong.  {\it Gaia}
DR2 provides no parallax because the star is a resolved 0\farcs2
binary BEU~13 with an estimated period of $\sim$80 yr.

{\bf HIP 46926} (BD+16~1992, G0, 107 pc) is a triple system. The outer
33\arcsec ~pair  is physical (common  PM and parallax).   We discovered
the  double-lined   inner  subsystem  Aa,Ab  with   $P=3.1$  days  and
determined  its  orbit.  Comparison  of  the RV  amplitudes  with  the
estimated masses implies a low orbital inclination of 15\degr.

{\bf  HIP 47110} (HD~82939,  G5V, 39  pc) is  a young  multiple system
belonging  to the  $\beta$ Pictoris  moving  group \citep{Alonso2015}.
The outer  pair with  a large separation  of 162\arcsec  ~is physical,
with common  PM, RV, and  parallax; its estimated period  is $\sim$160
kyr.   The M0V secondary  component B  (HIP~47133) was  found to  be a
double-lined spectroscopic  binary by \citet{Schlieder2012},  but they
have not provided its orbit. We observed the component B for almost 16
years and derive here an  orbit with $P=4.39$ days.  The residuals are
quite large,  2.7 \kms,  partly because the  star is faint  and partly
because  of its  chromospheric activity.   The outstanding  feature of
this orbit is  its large eccentricity of 0.47.  This  pair is still in
the phase of tidal  orbit circularization.  However, more observations
are needed to confirm and improve the orbit.

{\bf HIP  47899} (MCC~554, K4V,  79 pc)  is also a  double-lined binary  with a
period of 146  days and a small but  statistically significant orbital
eccentricity. The mass ratio is $q=0.94$.

{\bf HIP 48346 } (BD+38~2075, K8, 52 pc) is a double-lined binary with
$P=79$ d, rather large eccentricity $e=0.7$, and no other known visual
companions.

{\bf HIP 50156} (DK~Leo, M0.7V, 23  pc) is a triple system composed of
the 0\farcs1 visual binary A,B  with estimated period $\sim$3 yr, also
detected by astrometric  acceleration, and the 73-day single-lined
inner pair  Aa,Ab found here.  It is  a variable star of  BY Dra type.
The system belongs to the  $\beta$ Pictoris group according to several
authors  such as \citet{Schlieder2012,Alonso2015,Messina2017}  and has
an extensive bibliography.

{\bf HIP  50271 } (BD+26~2062, G0, 173 pc) is  a distant (869\arcsec) optical
companion to the  nearby (37 pc) G0V star  HIP~50355. Our observations
lead to the single-lined orbit with $P=47.5$ days.

{\bf HIP 52021 } (BD$-$05~3108, K8, 38 pc) has a variable RV, but our data do
not suffice to find its orbit.

{\bf HIP 54002 }  (AB Crt, K3V, 31 pc) is a  BY Dra type variable star
with variable RV, but no spectroscopic orbit yet.

{\bf HIP 54094}  (BD+54~1411,  unknown spectral type,  47 pc) has  the first
deviant  RV   measurement  that  suggests   its  variability.  Further
monitoring is needed.

{\bf HIP  56229} (  BD+41~2201, M0, 44  pc) is a  double-lined binary.
The 186-day  spectroscopic orbit derived  here from only eight  RVs is
not  very   secure.   The  star  is  an   X-ray  source.   Astrometric
acceleration was detected.

{\bf  HIP  57058}  (GJ~435.1,  K4V,   31  pc)  is  a  spectroscopic  and
acceleration  binary for  which we  derive a  preliminary single-lined
orbit   with  $P=726$\,d   by  combining   our  RVs   with   those  of
\citet{Halbwachs2018}. These  authors found that  the correlation dips
are double and  measured the RVs of the  secondary component. The VUES
profiles  are  double  as  well,  we  also measure  the  RVs  of  both
components.   This means  that the  CORAVEL RVs  derived by  fitting a
single component are biased both in our data and in those of Halbwachs
et al.  when the RV difference is small. We use CORAVEL RVs with a low
weight and disregard some of them.  The period and estimated masses of
0.69 \msun correspond  to a semimajor axis of  57\,mas.  This star was
resolved  twice   in  2018  by  speckle   inetrferometry  at the Southern
Astrophysical  Research Telescope,  SOAR, at  similar  separations and
shows a fast orbital motion,  in qualitative agreement with the 2-year
period (Tokovinin et al. 2019, in preparation).

The RVs of  the secondary component do not vary  in anti-phase with the
primary and, therefore, cannot be used to derive a double-lined orbit.
Hypothetically, this  is a triple system  where the outer  orbit has a
small  inclination (hence  small  $K_1$) and  its secondary  component
contains  a  short-period  subsystem.  Spectroscopic  monitoring  with
higher resolution  and future astrometric  orbits from {\it  Gaia} and
SOAR will clarify the architecture of this low-mass system.

{\bf BD+44  2120} (F5, 345 pc)  is a distant triple  system. The outer
6\farcs8 pair  ES~123 is physical,  based on the common  distances and
RVs  (the  PMs  are very  small).   Its  secondary  component B  is  a
single-lined binary with $P=2.2$ days.  We also observed the component
C at 42\arcsec  ~separation.  Its RV, as well as  PM, are distinct, so
the star C is unrelated (optical).

{\bf  GQ Leo}  (TYC~870-798-1,  K5Ve,  61 pc)  has  a slowly  variable
RV. Its mean  value of $-12.8$ \kms differs from  $-15.7 \pm 1.8$ \kms
measured by  {\it Gaia} and from  3 RVs around $-11$  \kms reported by
\citet{Griffin2005}. An orbit  with $P \approx 572$ d  and $K_1 = 2.8$
\kms can  be fitted to all  RVs. However, alternative  periods are not
excluded,  so we refrain  from publishing  this tentative  orbit.  The
object is  a 0\farcs25  pair MET~57Aa,Ab with  an estimated  period of
$\sim$45 years,  so some RV variation  could be caused  by this binary.
WDS lists another companion at 9\farcs5, which is optical according to
the {\it Gaia} astrometry.

{\bf HIP 57857} (G148-14, K0V, 54  pc) has a variable RV, but our data
are insufficient for  orbit determination. Astrometric acceleration is
evident  from  comparison  of  the  average PM  of  $(-339.3,  -44.4)$
mas~yr$^{-1}$ deduced from the  difference of {\it Hipparcos} and {\it
  Gaia}  positions  with the  ``instantaneous''  PM  measured by  {\it
  Gaia}:  $\Delta \mu  = (+18.2,  -6.1)$ mas~yr$^{-1}$.   The 4\farcs3
pair LEP~46  is physical according to  {\it Gaia} PM  and parallax, so
this is a triple system. The component  B has an RV of $-14.3 \pm 3.3$
(Gaia) \kms and its PM matches the average PM of A.

{\bf  HIP 57949}  (MCC~622, M0.5Ve,  31 pc)  also has  a  variable RV,
likely with a long period.

{\bf  HIP 59000}  (HD 105065,  K5V, 23  pc) is  a triple  or quadruple
system. The outer 183\arcsec ~binary (estimated period $\sim$300
kyr) is physical. The main  star is an acceleration binary. Our single
RV measurement, $-8.2$ \kms, differs  from the {\it Gaia} RV of $-29.2
\pm  13.6$  \kms.   The  large   error  of  the  latter  also  signals
variability.    Two   mutually  discordant   RVs   were  measured   by
\citet{Maldonado2010}.  It is not  clear whether the spectroscopic and
astrometric subsystems are same or distinct.

{\bf HIP 60433} (BD+21~2415, K4V, 40 pc) and {\bf HIP 60448} (MCC~654,
K5V, 30 pc) both have variable  RVs, but no orbits were computed.  The
first is also an acceleration binary.   Both stars have 18 years of RV
coverage.

{\bf HIP  61436} (GJ~9412, K5V,  30 pc) is a  double-lined binary with
$P=299$ d.

{\bf  HIP  62505}  (HD~111312, K2V,  26  pc)  is  a close  visual  and
spectroscopic binary  WSI~74 with  a period of  2.66 yr.  Its combined
orbit was determined by  \citet{Tok2017c}. Another visual companion at
2\farcs7 was measured  by {\it Hipparcos}, but never  confirmed; it is
not spotted by {\it Gaia} and therefore remains questionable.

{\bf HIP 62755} was originally  mis-identified with a nearby K6V dwarf
MCC  679.  The {\it  Gaia}  parallax of  $0.81  \pm  0.03$ mas  places
HIP~62755 among  giants.  We determined  an orbit with  $P=6.12$ years
from 11 RVs covering 8.2 years.

{\bf HIP 63253} (GJ~490, M0V, 21 pc) has an RV trend, so the period is
longer than 7 years. This is a 2+2 quadruple system. The outer 16\arcsec
~pair has a period of  $\sim$6 kyr. Both components were resolved into
0\farcs1 pairs with estimated periods  of a few years. The observed RV
variation is most likely related to the subsystem Aa,Ab.

{\bf HIP  63816} (GJ~497,  M0V, 16 pc)  is a 1\farcs6  visual binary
WOR~23. We find its RV constant over a time span of 7 years.

{\bf HIP 63942} (BD+21~2486, K4V, 19 pc) has a constant RV.  This is a
visual  binary  HU~739 with  an  orbital period  of  431  years and  a
semimajor axis of 2\farcs65.

{\bf HIP 65012} (GJ~507B, M3V, 14  pc) is the secondary component B of
HIP~65011  (at  17\farcs8  distance)  which  itself is  a  visual  and
spectroscopic  pair with $P=200$  d.  We  measured RV(B)  once, $-5.2$
\kms; {\it Gaia} measured it at $-9.1 \pm 0.6$ \kms.

{\bf  HIP  65026}  (HD~115953,  K0,  9  pc)  is  a  remarkable  triple
system.  The outer 1\farcs5  binary HU~644  has a  good-quality visual
orbit  with a  period of  49 yr.   The inner  subsystem Aa,Ab  is also
resolved as  CHR~193 at 0\farcs1. Here we  determined its single-lined
orbit  with a period  of 447  d.  A  preliminary period  of 450  d was
announced  by  \citet{Beuzit2004}.  A preliminary combined orbit of
the inner subsystem shows that the mutual inclination in this triple
system is small.

{\bf  HIP  65327}  (HD~238224,  K5V,  24  pc)  was  resolved  by  {\it
  Hipparcos}  at 0\farcs3  separation and  $\Delta m  = 2.3$  mag.  We
computed its single-lined orbit  with $P=12.4$ years.  However, the RV
amplitude  is  likely  reduced  by   line  blending.   A  combined
visual/spectroscopic orbit can be  computed now. WDS also mentions the
wide CPM  pair SHY~67 with  a separation of  9\fdg4, too wide to  be a
bound binary.

{\bf GJ 513} (M3V, 19 pc) has a slowly variable RV.

{\bf HIP 65887} (HD~117466, K0, 3 kpc) is a distant giant for which we
provide  a single-lined orbit  with a  3.3-year period.  Its semimajor
axis should be  11 mas, and, indeed, the  astrometric acceleration was
detected by {\it Hipparcos}.

{\bf HIP 66290 } (HD 118244, F5V, 38 pc) is a single-lined binary with
a period of  5.4 years, as well as acceleration  binary.  The orbit is
determined  from  36 RVs  measured  during 23  years.

{\bf HIP  67086} (K5,  46 pc)  is a 0\farcs6  binary resolved  by {\it
  Hipparcos}   (estimated   period    $\sim$100   yr)   containing   a
spectroscopic subsystem.  We measured RVs of both components with VUES
and determined  the orbit of the  secondary subsystem Ba,Bb  with $P =
41$ d.

{\bf BD+26 2498}  is a G5 giant with the DR2 parallax of $1.40 \pm 0.03$
mas. Our 31 RVs measured during 18 years securely define the
spectroscopic orbit with a period of 4.2 years. This orbit corresponds
to the minimum secondary mass of 1 \msun. The secondary component
could be a compact remnant. {\it Gaia} is expected to detect
acceleration  or deliver a full astrometric orbit.

{\bf HIP  67808} (BD+13~2721,  K7V, 22 pc)  is a 0\farcs2  visual (and
acceleration)  binary  with  an  estimated period  of  $\sim$10  years
discovered by  \citet{Beuzit2004}.  Our 6 RVs measured  during 7 years
are probably constant, with one measure deviating from the rest.

{\bf BD+19 2735} (K2, 37 pc) is a single-lined binary with a period of
almost 10 years; 1.5 orbital cycles of its eccentric orbit  are
covered.  Rotational modulation was measured by \citet{Kiraga2012}.

{\bf HIP 68801 }  (HD 123034, G5, 51 pc) is a double-lined binary with a
circular 2.8-day orbit and a mass ratio $q=0.98$ (a twin). \citet{N04}
discovered the RV variability but  provided no  orbit.

{\bf HIP  69549 } (HD 124605, G0,  85 pc) is a  double-lined pair with
$P=6$ d  and nearly equal components, $q=  0.98$.  The interferometric
pair 0\farcs08  TOK~723 with $\Delta  I =2.4$ mag  remains unconfirmed
and  could be  spurious;  its  parameters  are similar  to the  optical
ghosts reported in the discovery  paper by \citet{SAM17}. If it were a
triple system, the outer period would be around $\sim$10 yr.  However,
we do not see any  modulation of the center-of-mass velocity during 19
years covered by our data.

{\bf HIP 72508} (HR  5537, F5IV, 52 pc) has double lines,  but the period is
not known yet. The WDS companion B at 15\arcsec ~is optical according
to its {\it Gaia} astrometry. However, {\it Gaia} detected another faint ($G
= 17.42$ mag) star at 9\farcs2 separation with similar parallax and
PM, so this system is at least triple.

{\bf BD+49 2364}  is a giant, according to the  {\it Gaia} parallax of
$0.97 \pm 0.03$ mas.  Its RV shows  a trend and has changed by 5 \kms.
The orbital  period is  longer than  the time span  of our  data, 29.5
years.

{\bf HIP  76941} (MCC~316, K5V, 50  pc) is a  single-lined binary with
$P=267$ d.

{\bf  HIP 77141} (BD+36~2641,  K4/5V, 54\,pc)  has a  well-defined orbit
with $P=17.3$  d and  a very large  (for this period)  eccentricity of
$0.84\pm  0.15$.  Better  coverage near  periastron of  this  orbit is
needed to constrain the eccentricity.

{\bf HIP 78158}  (K5V, 52 pc) is a triple system  that consists of the
wide  187\arcsec ~physical  pair  A,B (LDS~983)  and the  single-lined
spectroscopic subsystem  Aa,Ab discovered here. Its  orbital period is
322 d. In  some spectra we noted secondary dips, but  their RVs do not
match the orbit.

{\bf HIP 79796} (BD+55~1823, CR~Dra, M5.6V, 20 pc) is a low-mass flare
star and an interferometric binary BLA~3, for which an orbit with 4.04
yr period  was computed by  \citet{Tam2008}. We see double  lines, but
their RVs do  not match the visual orbit. A tentative  RV curve with a
period    of   1.57    yr   is    plotted    in   Fig.~\ref{fig:orb3}.
\citet{Shkolnik2010} obtained two  double-lined spectra and determined
that the period is less than  530 d.  More work is needed to reconcile
RVs with position measurements and, hopefully, to compute the combined
orbit.

{\bf HIP  80751} (BD+24~3014, K5V, 32  pc) has a  variable RV measured
during 18  years.  We determined  a tentative orbit with  $P=14.4$ yr,
but its  confirmation by  new measurements and  a better  coverage are
needed. Astrometric acceleration was detected.

{\bf  BD+52 1968}  (K8,  44 pc)  is  a 5\farcs6  visual binary  ES~968
(estimated period $\sim$3 kyr). The RV of the component A, measured 13
times with VUES, is certainly variable  with a long period.  The RV of
B was measured only on one night and agrees with the mean RV of A.

{\bf HIP 82506 } (HD~152342, F4III,  67 pc) has a variable RV, but not
enough data  for orbit calculation.  Is is also an  astrometric binary.

{\bf BD+61  1678C} (GJ~685, M1V, 14  pc) is the  distant (738\arcsec)
component to the visual pair A,B  (GJ 684, HIP 86036, G0V) which has a
period  of  76 yr  and  the  corresponding single-lined  spectroscopic
orbit.  We monitored  RV of  the component C and found it constant,
agreeing  with  the RV  of  A.  The  same  conclusion  was reached  by
\citet{Tok1992}.

{\bf HIP 90274 } (HD 170527, K0, 175 pc) is a giant observed during 16
years. The spectrum has blended  double lines, but no orbit is derived
yet.

{\bf HIP 91043 } (HD~171488, V889  Her, G2V, 35 pc) has a double-lined
spectrum  and  no spectroscopic  orbit  despite  our 39  observations,
mostly  with unresolved  CORAVEL  dips. Several  faint companions  are
listed in the WDS, but none of those is confirmed as physical.

{\bf  HIP 92952} (G~229-18,  M0V, 46  pc) is  a quadruple  system. The
outer  pair A,B  (GIC~154)  has  a separation  of  119\arcsec ~and  an
estimated  period of  $\sim$300 kyr.  The  component A  is a  0\farcs4
visual  binary Aa,Ab  resolved  by {\it  Hipparcos}  and not  measured
since; its estimated period is $\sim$70 years.  We see double lines in
the spectrum.  Stationary lines  correspond to the visual secondary Ab
and  the moving  lines to  the primary  Aa, which  is  a spectroscopic
binary  Aa1,Aa2  with  a period  of  8  d.   The eccentricity  of  the
spectroscopic  orbit is  small,  but statistically  significant, $e  =
0.044 \pm  0.007$.  The  RV of the  star Ab  is about $-15$  \kms. Its
measurements are not accurate owing to blending with the lines of Aa1.

{\bf HIP 94557} (G~185-12, M4.5V,  18 pc) is the brighter component of
the wide  visual binary WDS J19147+1918 (LDS~2020,  41\arcsec).  Our 3
RVs show variability.   \citet{Shkolnik2012} measured a very different
RV of $-80.9$ \kms, quoted in Simbad.

{\bf  HIP 94622}  (GJ 751,  M0, 29  pc) has  a variable  RV,  and some
spectra  have double  lines.  The  RV  variability was  also noted  by
\citet{Tok1992}.   This  is a  {\it  Hipparcos}  binary with  0\farcs2
separation and  an estimated period of  $\sim$10 yr.  It  is not clear
whether the  visual and spectroscopic  systems correspond to  the same
pair or, alternatively, if it is a triple system.

{\bf BD+77 767}   (K8, 41 pc) is a single-lined binary with $P=17.3$ d.

{\bf HIP 99969}  (BD+06~4489, K4V, 44 pc) has double lines, but we are not
yet able to determine its orbit.

{\bf HIP 101941}  (HD~196928, K4III, 380  pc) is a  single-lined binary with
$P=2.8$ yr, as well as an astrometric binary.

{\bf  HIP 102300}  (M0Ve,  21 pc)  has a  variable  RV. The
period is longer than 8 yr covered by our data.

{\bf HIP  102320} (HD~335007,  K5, 42 pc)  belongs to a  triple system
where the  outer 4\farcs4  pair ES~366  has a period  of the  order of
$\sim$2  kyr and  $\Delta  m  = 3$  mag.  The main  component  A is  a
double-lined  twin binary  with $P=21$  d and  equal components,  $q =
0.97$.   The orbital inclination  of the  spectroscopic pair  is about
55\degr.

{\bf HIP 102718 } (BD+04~4551,  F7Vw, 103 pc) is the primary component
of    WDS    J20488+0512, a 6\farcs6 pair.    Its    RV   is    likely
constant.

{\bf HIP 103375  } (HD~235405, G0, 132 pc)  belongs to WDS J20566+5250
(ADS 14465, A~1437, separation 1\farcs3). Its RV is likely constant.

{\bf HIP  104994} (BD+28~4035, G5,  143 pc) is a  triple-lined system.
The outer 0\farcs19  binary was first resolved by  {\it Hipparcos} and
not measured since;  its period is $\sim$100 yr.   Stationary lines in
the spectrum belong  to the visual secondary B with  RV of $-54$ \kms.
The  pair of  moving lines  corresponds  to the  subsystem Aa,Ab  with
$P=52$ d  and systemic velocity  of $-46.8$ \kms.  Its  inclination is
about 52\degr.

{\bf HIP 105504}  (HD~358435, K7, 42  pc) definitely has a  variable RV, but
not enough data for an orbit. Its astrometric acceleration was
detected.

{\bf BD+47 3439} (K0, 115 pc) is an evolved star in the visual triple system
ADS~15052, with separations of 1\farcs1 and 0\farcs2. The spectra have
double lines, but we could not  yet figure out the orbit. Either there
is an additional spectroscopic subsystem, or the inner binary is going
through the periastron of its eccentric orbit (its estimated period is
$\sim$80 years). The fact that {\it Gaia} DR2 measured the parallax
favors the second option because partially resolved sources do not
have parallaxes in DR2.

{\bf HIP 110291 } (HD 212029, G0, 62 pc) is a single-lined binary with
a period of  2.1 yr.  A similar period was  determined by D. Latham
(2012,  private  communication),  while  two astrometric  orbits  with
periods of 2.07 and 2.17  yr and large eccentricities were computed
by \citet{Goldin2006}. WDS lists several optical companions.

{\bf HIP  110526} (GJ~826, M3.0V,  15 pc)  has a  constant RV  over eight  years  of our
monitoring. It is a visual binary WOR~11 with $P=130$ yr and semimajor
axis 1\farcs61.

{\bf HIP 110978 } (HD 213054, K2III,  800 pc) may have a mild RV trend
over 34 days. It has astrometric acceleration.

{\bf HIP  111685} (BD+38~4818, M0Ve, 24  pc) is the  main component of
the triple  system. The faint ($V=21.2$  mag) outer component  C is at
33\farcs4 distance  from the main  star, with common PM  and parallax.
The  inner  pair  A,B,  first  resolved  by  {\it  Hipparcos},  has  a
well-defined visual  orbit with $P=16$ yr.  We fitted our  RVs and all
available positional  measurements to a  combined orbit and  give here
its spectroscopic elements.

{\bf HIP  111942} (GJ~870, K8V,  31 pc) is a  single-lined binary with
$P=375$ days. The orbit derived from our 20 RVs is very similar to the
orbit published independently by \citet{Halbwachs2018} and based on 25
CORAVEL  RVs. Here  we list  orbital elements  fitted to  the combined
data, with global rms residual of 0.32 \kms.

{\bf HIP 112040}  (BD+18~5029, M0V, 31 pc) has a variable RV  with a period longer
than 15 yr, as well as astrometric acceleration.

{\bf  HIP 112268}  (BD+16~4806, K6V, 50  pc)  has a  variable RV.  We computed  an
uncertain orbit with  $P=4.35$ yr which is not  publishable until a
better coverage is obtained.

{\bf HIP 112523} (MCC~851, K5V, 37 pc)  has a preliminary orbit with $P=8.9$ yr
derived from 13 RVs spanning the period of 15 yr. The eccentricity was
fixed in the orbit fitting. Astrometric acceleration was detected.

{\bf HIP 116003} (GJ~1284, M2Ve, 16 pc) is a flare star and an X-ray source. As
our two  RVs are  very different from  each other,  it must be  a fast
spectroscopic binary.  Variable RV was noted by \citet{Gizis2002}.

{\bf BD+66 1664}  (G5, 90 pc) is a single-lined binary with a circular 3-day
orbit. Fast synchronous rotation is responsible for its high chromospheric
activity.

{\bf HIP 118212} (GJ~913, K7V, 17  pc) is a single-lined binary with a
period  of 872  d (2.4  yr).  It  is also  an acceleration  binary and
suspected  non-single star  in {\it  Hipparcos}.   One interferometric
resolution  at 62\,mas  with  $\Delta I  =  1.4$ mag  was reported  by
\citet{Bag2007b}.   The astrometric  orbit by  \citet{Goldin2006} with
$P=885$ d and $e=0.56$ is similar to our spectroscopic orbit. However,
their revised parallax of 67\,mas is not confirmed by {\it Gaia}.

\subsection{The triple system HIP 13460}

 \begin{figure}
   \centering
   \includegraphics[width=\hsize]{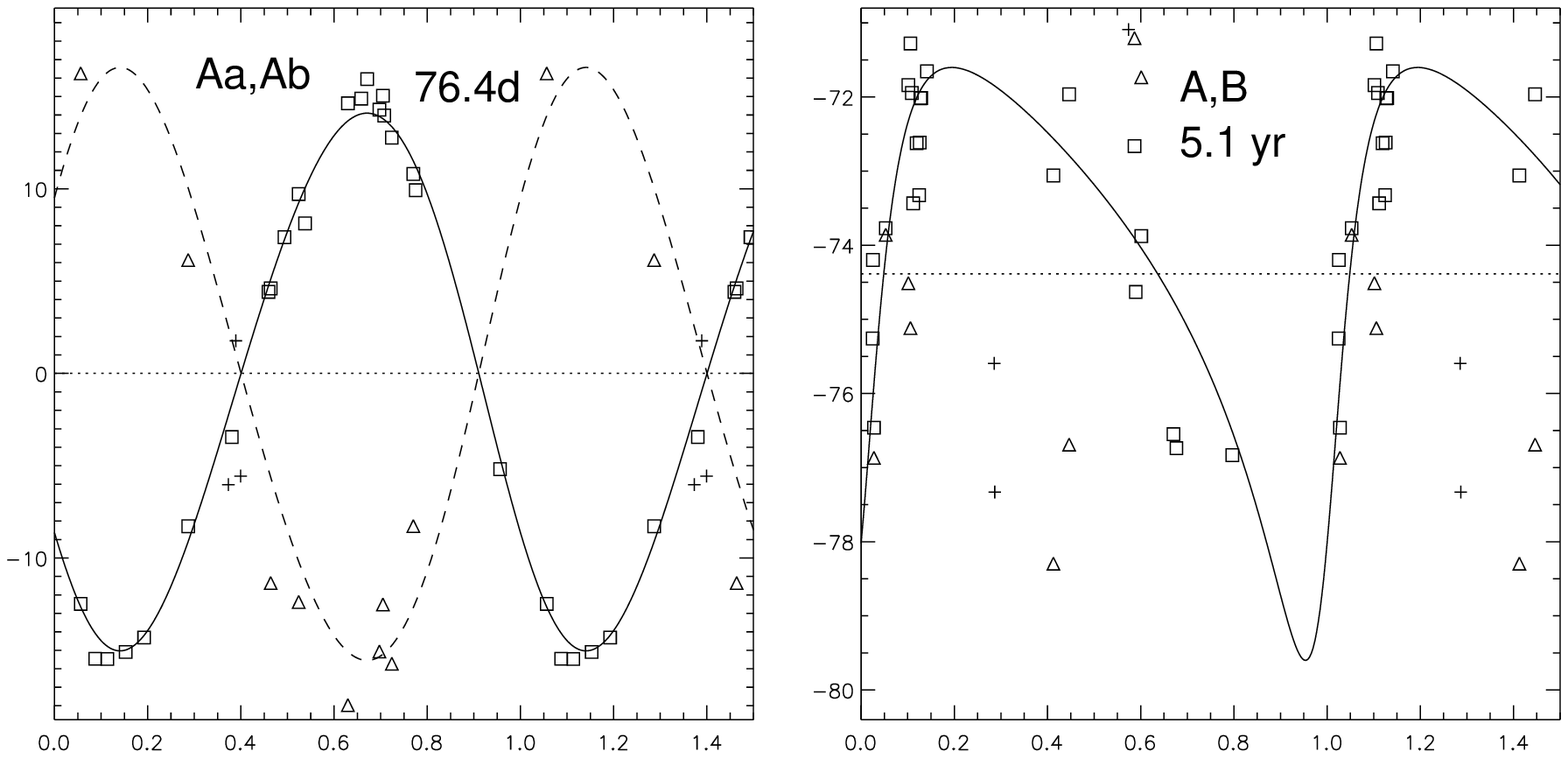}
      \caption{RV curves of HIP~13460. Left: inner subsystem, $P=76.4$
        d,  right: outer  subsystem,  $P=5.1$ yr.   RVs derived  from
      blended dips are plotted as  crosses and are given small weights
      in the orbit fit.
         \label{fig:triple} }
   \end{figure}

This star,  also known  as BD+60~585  and GJ~3185, is  a K3V  dwarf at
39\,pc from the  Sun. The spectrum is double-lined,  and the period of
76~d is  readily found. However,  all 30 RVs  cannot be fitted  by the
common elements, leaving residuals of 1.6  and 2.0 \kms for Aa and Ab,
respectively.  Individual  fits to the  RVs measured with  CORAVEL and
VUES are better, but result in slightly different elements.

All RVs  can be modeled better  by assuming that the  pair Aa,Ab moves
slowly  on  an  outer  orbit.  Astrometric  acceleration  reported  by
\citet{MK05}  and confirmed  by {\it  Gaia} supports  the triple-star
hypothesis.  We fitted the long-period orbit with $K_1 = 4$ \kms using
{\tt  orbit3.pro} (Fig.~\ref{fig:triple}).  The rms  residuals  to the
triple-star solution  are 0.88 \kms for  Aa and 3.23 \kms  for Ab. The
component Ab may contain a short-period subsystem.

The masses of Aa and Ab estimated from the absolute magnitudes and the
mass ratio are 0.78 and 0.71 \msun. Comparison with the minimum masses
leads to the inclination $i_{\rm  Aa,Ab} = 32^\circ$. The minimum mass
of the tertiary component B is 0.29 \msun. Although the semimajor axis
of the outer orbit, computed from the period and mass sum, is 92\,mas,
there is little  hope of resolving A,B directly  owing to the expected
faintness of B. On the  other hand, {\it Gaia} can provide astrometric
orbits of both inner and outer systems.  To do so, an initial guess of
the orbital  periods and other  parameters will likely be  needed, and
our work  provides these  parameters. It is  unlikely that  {\it Gaia}
astrometry of this complex system  can be interpreted correctly by its
pipeline alone without additional inputs.

\section{Summary}
\label{sec:sum}

The main results of this work are:

\begin{itemize}
\item
A large set of RV measurements spanning three decades.
\item
Determination of 57 spectroscopic orbits, 53 of those for the first time.

\item
Discovery of 20 new nearby hierarchical systems.

\item
Discovery of  interesting stellar systems.  For example,  in the young
triple HIP~47110 the inner orbit with $P=4.4$ d and $e= 0.47$ is still
circularizing, apparently.

\end{itemize}

\begin{figure}
   \centering
   \includegraphics[width=\hsize]{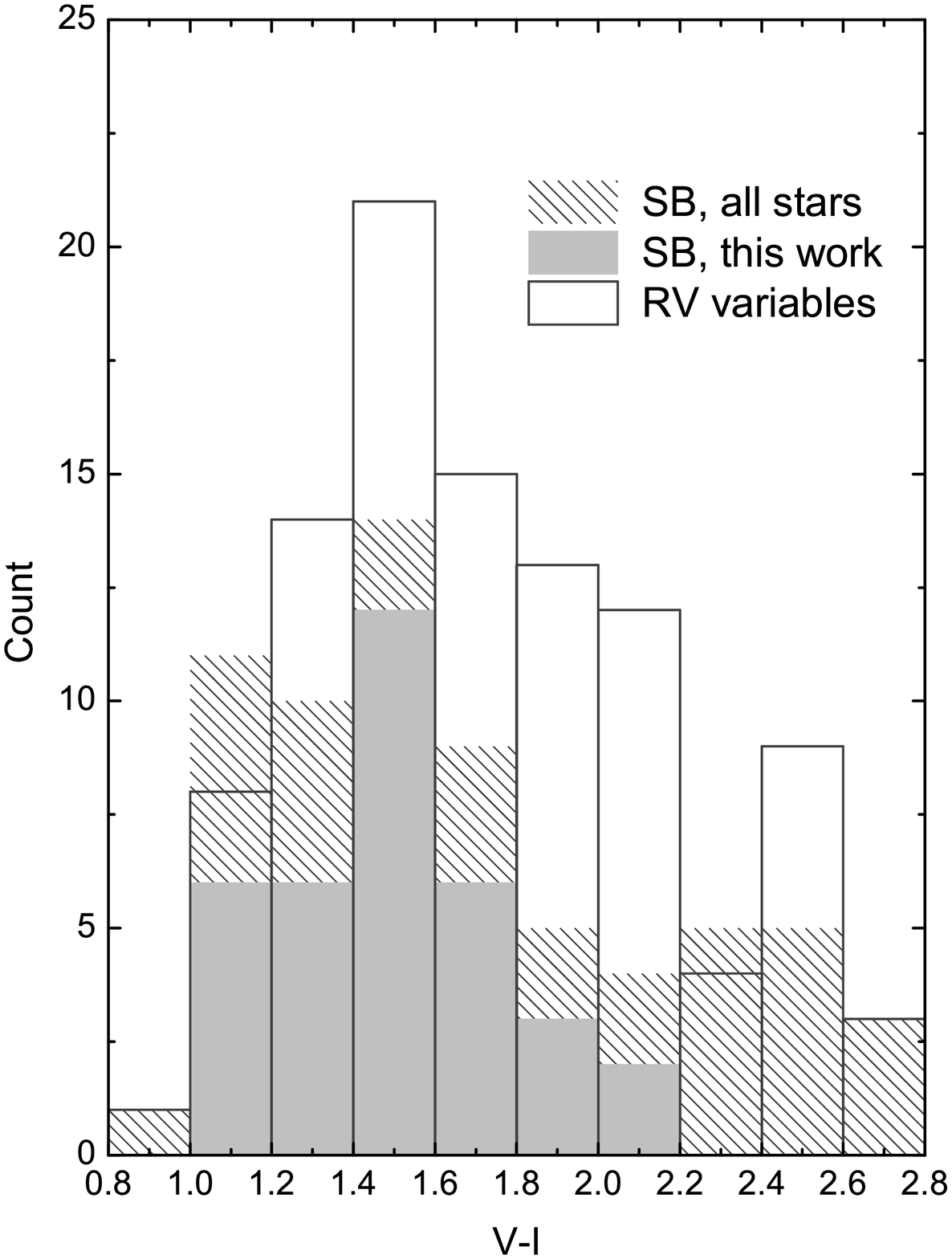}
      \caption{Histogram of the number of spectroscopic binaries and
        stars with variable RV in the sample of K- and M-dwarfs of \citet{S16}.
}
         \label{fig:orbits}
\end{figure}

Most orbits  presented here refer to  nearby K- and  M-type dwarfs and
result from the long-term RV monitoring. Our sample includes 857 stars
from   the  McCormick   catalog  and   188  stars   from   the  Gliese
catalog. Observational  data on  this sample are  given in  Table~4 of
\citet{S16}. A  total of 67  spectroscopic orbits are known  for these
stars, including 35  determined here. In addition, there  are 70 stars
with variable  RV without orbital  elements identified from  our data,
from the literature \citep[e.g.][]{Halbwachs2018}, or by comparing our
RVs with those from {\it Gaia} DR2. The latter group counts 30 objects
where the RV  difference exceeds 2.5 \kms (3$\sigma$).  For stars with
constant RV, the mean RV difference between our CORAVEL and {\it Gaia}
RVs  is   0.21  \kms  with  the   rms  scatter  of   $\sigma  =  0.74$
\kms. Figure~\ref{fig:orbits} illustrates  the contribution of this work
to the census of spectroscopic binaries among K- and M-dwarfs.

We continue observations of the  identified RV variables with the VUES
spectrograph  in order to  detect double-lined  spectroscopic binaries
among them  and to  calculate their orbital  parameters. Such  type of
stars may not be recognized by {\it Gaia} due to rather low resolution
of its spectrometer.

\begin{acknowledgements}

It is a pleasure to thank an anonymous referee for careful reading 
and valuable comments. We  used the  Simbad  service  operated by  the  
Centre des  Donn\'ees Stellaires (Strasbourg, France).  This work also
made use of data from the     European      Space     Agency     (ESA)      
mission     {\it  Gaia}\footnote{https://www.cosmos.esa.int/gaia},  
processed  by  the {\it   Gaia}   Data   Processing   and  Analysis   Consortium   
(DPAC, https://www.cosmos.esa.int/web/gaia/dpac/consortium).  Funding for the
DPAC  has been provided  by national  institutions, in  particular the
institutions participating in the {\it Gaia} Multilateral Agreement.

\end{acknowledgements}

%
%

\bibliographystyle{aa} 

\bibliography{rvvar_2a}



%

\longtab{
\begin{longtable}{l  llll lll l l}
\caption{Spectroscopic orbits \label{tab:sborb}   } \\
\hline\hline
Name & $P$ & $T$ & $e$ & $\omega_{\rm A}$ & $K_1$ & $K_2$ &  $\gamma$ & rms$_{1,2}$ & $M_{1,2} \sin^3 i$ \\
  & (d) & (+24\,00000) & (deg) & (km~s$^{-1}$) & (km~s$^{-1}$) &
(km~s$^{-1}$) & (km~s$^{-1}$) & & (${\cal M}_\odot$) \\
\hline
\endfirsthead
\caption{continued.}\\
\hline\hline
Name & $P$ & $T$ & $e$ & $\omega_{\rm A}$ & $K_1$ & $K_2$ &  $\gamma$ & rms$_{1,2}$ & $M_{1,2} \sin^3 i$ \\
  & (d) & (+24\,00000) & (deg) & (km~s$^{-1}$) & (km~s$^{-1}$) &
(km~s$^{-1}$) & (km~s$^{-1}$) & & (${\cal M}_\odot$) \\
\hline
\endhead
\hline
\endfoot
BD+33 4827D & 67.9618 & 51794.504 & 0.197 & 280.4 & 10.263 & \ldots & $-$23.478&  0.58   & $>$0.2  \\
         & $\pm$0.0035 & $\pm$0.633 & $\pm$0.010 & $\pm$3.6 & $\pm$0.125 & \ldots & $\pm$0.080&   \ldots   &  \ldots \\
HIP 3428 & 97.250 & 58183.305 & 0.391 & 21.6 & 25.244 & 26.490 & $-$6.648&   0.52  &  0.56 \\
         & $\pm$0.015 & $\pm$0.266 & $\pm$0.005 & $\pm$1.1 & $\pm$0.197 & $\pm$0.211 & $\pm$0.072& 0.46  & 0.53 \\
HD 8691 & 581.17 & 55965.5 & 0.544 & 313.5 & 2.366 & \ldots & $-$36.183&  0.54   & $>$0.08  \\
         & $\pm$1.21 & $\pm$5.1 & $\pm$0.036 & $\pm$5.5 & $\pm$0.110 & \ldots & $\pm$0.065&   \ldots  & \ldots  \\
HIP 9867 & 897.0 & 56004.0 & 0.0 & 0.0 & 8.057 & 11.347 & 61.419  &  0.38   & 0.40 \\
         & $\pm$2.3 & $\pm$5.8 & fixed & fixed & $\pm$0.114 & $\pm$0.184 & $\pm$0.079& 0.37    & 0.28  \\
HIP 10258 & 5.88554 & 54206.1875 & 0.0   & 0.0 & 66.011 & 64.231 & $-$59.148&  0.92   &  0.66 \\
         & $\pm$0.00001 & $\pm$0.0022 & fixed & fixed & $\pm$0.176 & $\pm$0.189 & $\pm$0.085& 0.88    & 0.68 \\
HIP 13460 Aa,Ab & 76.433  & 58410.20 & 0.057      & 124.1     & 14.565      & 16.060     &  \ldots & 0.88    & 0.12 \\
         & $\pm$0.014  & $\pm$3.20   & $\pm$0.016 & $\pm$14.2 & $\pm$0.254  & $\pm$1.508 &  \ldots & 3.23    & 0.11 \\
HIP 13460 AB & 1862: & 55433    & 0.502      & 232.8    & 4.00      & \ldots & $-$74.386  & \ldots    & $>$0.29  \\
         & $\pm$25  & $\pm$149 & $\pm$0.186 & $\pm$7.4 & $\pm$2.04 & \ldots & $\pm$0.577 &  \ldots   & \ldots  \\
HIP 14478 & 1325.2 & 57011.8 & 0.369 & 142.4 & 5.814 & \ldots & $-$50.871&  0.22   & $>$0.34  \\
         & $\pm$7.8 & $\pm$16.5 & $\pm$0.033 & $\pm$5.8 & $\pm$0.162 & \ldots & $\pm$0.183&   \ldots   &  \ldots \\
HIP 14864 & 59.428 & 56470.94 & 0.642           & 130.3   & 27.626    & 31.667      & $-$31.907&  0.37   & 0.33 \\
         & $\pm$0.005 & $\pm$0.07 & $\pm$0.006 & $\pm$0.8 & $\pm$0.256 & $\pm$1.020 & $\pm$0.138&  1.15   & 0.29 \\
HIP 17102 & 183.113  & 57305.62   & 0.6115         & 350.09    & 25.134     & 24.121    &  4.273     & 0.42    & 0.56  \\
         & $\pm$0.019  & $\pm$0.17    & $\pm$0.0020& $\pm$0.49 & $\pm$0.124 &$\pm$0.128 & $\pm$0.073 & 0.41    & 0.58 \\
HIP 18448 & 1538.9 & 54793.6 & 0.478 & 210.6 & 6.469 & \ldots & 16.657&   0.34  & $>$0.38 \\
         & $\pm$2.6 & $\pm$5.9 & $\pm$0.018 & $\pm$2.3 & $\pm$0.117 & \ldots & $\pm$0.088&   \ldots   &  \ldots \\
HIP 19915 & 7.30032 & 50087.8789 & 0.0 & 0.0 & 27.326 & 30.620 & $-$8.157&  1.25   & 0.078 \\
         & $\pm$0.00003 & $\pm$0.0092 & fixed & fixed & $\pm$0.160 & $\pm$0.182 & $\pm$0.107& 1.44    & 0.069 \\
HD 279846 & 15.4860 & 55271.598 & 0.039 & 245.1 & 44.723 & 47.041 & 12.429&  0.78   &  0.63 \\
         & $\pm$0.0001 & $\pm$0.189 & $\pm$0.003 & $\pm$4.5 & $\pm$0.217 & $\pm$0.225 & $\pm$0.075&  1.09   & 0.60 \\
HIP 20709 & 140.748 & 56222.61 &  0.557       & 280.0     & 23.200    & 31.750      & 44.060 &  0.19   &  0.84 \\
         & $\pm$0.020 & $\pm$0.34 & $\pm$0.007 & $\pm$0.6 & $\pm$0.086 & $\pm$0.332 & $\pm$0.067 &  0.19   & 0.60 \\
HIP 21710 & 610.43 & 56635.1 & 0.415         & 269.9    & 4.596 & \ldots & $-$27.124  &  0.75   & $>$0.17  \\
         & $\pm$0.15 & $\pm$4.2 & $\pm$0.018 & $\pm$3.4 & $\pm$0.099 & \ldots & $\pm$0.067&   \ldots   &  \ldots \\
HIP 28663 & 8.5542 & 52158.14 & 0.079 & 255.3 & 30.129 & \ldots & $-$25.105&  0.68   &  $>$0.35 \\
         & $\pm$0.0001 & $\pm$0.12 & $\pm$0.006 & $\pm$5.0 & $\pm$0.229 & \ldots & $\pm$0.136&   \ldots   &  \ldots \\
HIP 34341 & 875.37 & 55875.7 & 0.507 & 328.7 & 4.561 & \ldots & $-$17.353&  0.45   &  $>$0.20\\
         & $\pm$1.69 & $\pm$8.6 & $\pm$0.026 & $\pm$5.4 & $\pm$0.284 & \ldots & $\pm$0.106&   \ldots   & \ldots  \\
HIP 35706 & 4741 & 56641 & 0.337 & 266.5 & 5.639 & \ldots & $-$29.953&  0.22   & $>$0.57  \\
         & $\pm$73 & $\pm$41 & $\pm$0.022 & $\pm$4.6 & $\pm$0.098 & \ldots & $\pm$0.125&   \ldots   &  \ldots \\
HIP 39681 & 1717.7 & 51745.8 & 0.508 & 76.3 & 8.077 & \ldots & $-$0.569& 0.47    & $>$0.51  \\
         & $\pm$0.7 & $\pm$3.7 & $\pm$0.009 & $\pm$1.5 & $\pm$0.116 & \ldots & $\pm$0.054&  \ldots    &  \ldots \\
HIP 40253& 897.83 & 54881 & 0.035 & 340 & 6.638 & \ldots & $-$3.255&  0.51   & $>$0.37  \\
         & $\pm$0.86 & $\pm$66 & $\pm$0.010 & $\pm$26.5 & $\pm$0.151 & \ldots & $\pm$0.060&   \ldots   & \ldots  \\
HD 71028 & 1129.5 & 53846.9 & 0.147 & 166.1 & 4.989 & \ldots & 34.147&  0.41   & $>$0.29  \\
         & $\pm$3.2 & $\pm$42.6 & $\pm$0.028 & $\pm$14.2 & $\pm$0.174 & \ldots & $\pm$0.086&   \ldots   &  \ldots \\
HIP 46383 & 8.490802 & 57396.839      & 0.0987     & 324.3     & 56.286 & 57.002 & $-$32.322        & 0.38     & 0.63 \\
         & $\pm$0.000003 & $\pm$0.013 & $\pm$0.0016 & $\pm$0.6 & $\pm$0.095 & $\pm$0.096 & $\pm$0.005& 0.43    & 0.63 \\
HIP 46926 & 3.10662 & 54604.7305 & 0.0 & 0.0 & 22.600 & 26.464 & $-$10.859&  0.59   & 0.020 \\
         & $\pm$0.00001   & $\pm$0.0025 & fixed & fixed & $\pm$0.127 & $\pm$0.185 & $\pm$0.065& 0.50    & 0.017 \\
HIP 47133 & 4.38804 & 58055.344 & 0.474 & 140.2 & 69.252 & 74.379 & $-$0.677&  2.71   &  0.47 \\
         & $\pm$0.00001 & $\pm$0.015 & $\pm$0.005 & $\pm$1.9 & $\pm$2.035 & $\pm$2.506 & $\pm$0.257&  2.79   &  0.44 \\
HIP 47899 & 146.309 & 56614.18 & 0.075 & 5.8 & 20.100 & 21.360 & $-$15.010&  0.66   & 0.55 \\
         & $\pm$0.020 & $\pm$1.50 & $\pm$0.005 & $\pm$3.6 & $\pm$0.131 & $\pm$0.144 & $\pm$0.071& 0.63    & 0.52 \\
HIP 48346  & 79.013 & 56464.52 & 0.695        & 187.9     & 26.408     & 36.202     & $-$14.648&  0.86   &  0.43 \\
         & $\pm$0.073 & $\pm$1.03 & $\pm$0.041 & $\pm$2.5 & $\pm$3.209 & $\pm$4.602 & $\pm$0.308&  2.82   & 0.31 \\
HIP 50156 & 73.225 & 56288.11 & 0.140 & 308.7 & 9.851 & \ldots & 8.741&   0.63  & $>$0.22  \\
         & $\pm$0.030 & $\pm$2.10 & $\pm$0.027 & $\pm$11.1 & $\pm$0.266 & \ldots & $\pm$0.222&  \ldots   & \ldots \\
HIP 50271  & 47.4632 & 49595.278 & 0.688 & 55.6 & 28.703 & \ldots & 4.288&  0.98   & $>$0.45  \\
         & $\pm$0.0003 & $\pm$0.028 & $\pm$0.003 & $\pm$0.6 & $\pm$0.156 & \ldots & $\pm$0.108&  \ldots   & \ldots \\
HIP 56229 & 186.3:  & 58096.59 & 0.549 & 345.7 & 22.183 & 23.033 & 12.015&  0.42   & 0.53 \\
         & $\pm$0.73 & $\pm$1.04 & $\pm$0.050 & $\pm$1.7 & $\pm$2.298 & $\pm$2.386 & $\pm$0.064&  0.33   & 0.51 \\
HIP 57058 &  725.9 & 57934.3         & 0.164 & 168.0    &   8.065 & \ldots & 19.708& 0.53    & $>$0.50  \\
         & $\pm$0.9 & $\pm$61.7 & $\pm$0.079 & $\pm$28.9 & $\pm$1.128 & \ldots & $\pm$0.603& \ldots    & \ldots \\
BD+44 2120B & 2.22596 & 55870.6875 & 0.0 & 0.0 & 46.215 & \ldots & $-$4.387&  1.15   & $>$0.35  \\
         & $\pm$0.00001  & $\pm$0.0038 & fixed & fixed & $\pm$0.322 & \ldots & $\pm$0.218& \ldots    &\ldots  \\
HIP 61436 & 299.55 & 56819.02 & 0.457 & 260.7 & 17.590 & 19.677 & $-$0.363&  0.48   & 0.49 \\
         & $\pm$0.18 & $\pm$1.24 & $\pm$0.013 & $\pm$1.9 & $\pm$0.294 &  $\pm$0.491 & $\pm$0.135& 0.96    & 0.46 \\
HIP 62755 & 2238 & 56347.9 & 0.140 & 27.3 & 8.027 & \ldots & $-$47.032& 0.11    & $>$0.69 \\
         & $\pm$27 & $\pm$127.6 & $\pm$0.024 & $\pm$20.9 & $\pm$0.419 & \ldots & $\pm$0.167&   \ldots  & \ldots \\
HIP 65026 & 446.87 & 57009.7      & 0.082     & 12.2     & 9.129      & \ldots & 0.109&   0.65  & 0.65: \\
         & $\pm$0.43 & $\pm$10.1 & $\pm$0.014 & $\pm$8.4 & $\pm$0.158 & \ldots & $\pm$0.104& \ldots  & $>$0.29   \\
HIP 65327 & 4354.0 & 56827.0 & 0.497          & 115.2    & 5.030      & \ldots & $-$8.528  &  0.21   & $>$0.45  \\
         & $\pm$13.8 & $\pm$55.8 & $\pm$0.051 & $\pm$8.7 & $\pm$0.557 & \ldots & $\pm$0.20 &  \ldots   & \ldots \\
HIP 65887 & 1200.0 & 53733.7 & 0.078 & 68.6 & 6.143 & \ldots & $-$3.416&  0.45   & $>$0.38  \\
         & $\pm$3.0 & $\pm$47.8 & $\pm$0.020 & $\pm$14.5 & $\pm$0.134 & \ldots & $\pm$0.074&  \ldots   & \ldots \\
HIP 66290 & 1967 & 51294.7 & 0.681 & 78.5 & 6.731 & \ldots & $-$14.044&  0.54   & $>$0.36  \\
         & $\pm$11 & $\pm$9.1 & $\pm$0.018 & $\pm$5.8 & $\pm$0.315 & \ldots & $\pm$0.139& \ldots    &\ldots  \\
HIP 67086 & 40.871    & 56774.02   & 0.318      & 208.9    & 28.116     & 33.602     & $-$27.226  &  0.55   & 0.46     \\
         & $\pm$0.007  & $\pm$0.34 & $\pm$0.024 & $\pm$2.8 & $\pm$0.564 & $\pm$1.070  & $\pm$0.255 &  0.54   & 0.39   \\
BD+26 2498 & 1536.2 & 53208.1 & 0.190 & 258.3 & 12.063 & \ldots & $-$13.641&  0.52   & $>$1.02  \\
         & $\pm$1.8 & $\pm$15.0 & $\pm$0.010 & $\pm$3.5 & $\pm$0.117 & \ldots & $\pm$0.100&   \ldots  &\ldots  \\
BD+19 2735 & 3616 & 55769.8 & 0.607 & 89.8 & 4.913 & \ldots & $-$9.535& 0.28    &  $>$0.34 \\
         & $\pm$69 & $\pm$33.2 & $\pm$0.176 & $\pm$7.9 & $\pm$1.566 & \ldots & $\pm$0.120&  \ldots   & \ldots \\
HIP 68801 & 2.84019 & 54536.9453 & 0.0 & 0.0 & 86.676 & 88.273 & $-$36.319& 1.10    & 0.79 \\
         & $\pm$0.00001 & $\pm$0.0009 & fixed & fixed & $\pm$0.173 & $\pm$0.222 & $\pm$0.099& 0.90    & 0.78 \\
HIP 69549 & 6.03150 & 54087.0820 & 0.0 & 0.0 & 64.754 & 65.831 & 8.578&  1.13   & 0.70 \\
         & $\pm$0.00001 & $\pm$0.0016 & fixed & fixed & $\pm$0.132 & $\pm$0.139 & $\pm$0.068& 1.36    & 0.69 \\
HIP 76941 & 267.06   & 57157.05  & 0.492      & 315.3    & 7.054     & \ldots & $-$12.497  &  0.38   & $>$0.23  \\
         & $\pm$0.31 & $\pm$1.36 & $\pm$0.012 & $\pm$1.7 & $\pm$0.122 & \ldots & $\pm$0.071&  \ldots   & \ldots \\
HIP 77141 & 17.3120 & 58299.184 & 0.836 & 14.2 & 50.441 & \ldots & $-$3.460&  0.28   & $>$0.34  \\
         & $\pm$0.0005 & $\pm$0.070 & $\pm$0.154 & $\pm$8.1 & $\pm$58.975 & \ldots & $\pm$2.167& \ldots    & \ldots \\
HIP 78158 & 322.04 & 57083.48 & 0.395 & 299.7 & 11.784 & \ldots & $-$44.839&  0.34   & $>$0.44 \\
         & $\pm$0.16 & $\pm$1.43 & $\pm$0.014 & $\pm$1.9 & $\pm$0.219 & \ldots     & $\pm$0.099&  \ldots   & \ldots \\
HIP 80751 & 5252 & 58096.9 & 0.188 & 64.5 & 3.031 & \ldots & $-$32.748&  0.23   &   $>$1.02  \\
         & $\pm$324 & $\pm$692.7 & $\pm$0.069 & $\pm$55.1 & $\pm$0.748 & \ldots & $\pm$0.582&   \ldots  & \ldots \\
HIP 92952 & 7.9461 & 57609.65 & 0.044 & 291.9 & 17.290 & \ldots & $-$18.380&  0.40   & $>$0.18  \\
         & $\pm$0.0002 & $\pm$0.28 & $\pm$0.007 & $\pm$12.8 & $\pm$0.148 & \ldots & $\pm$0.133& \ldots    & \ldots \\
BD+77 767 & 17.3231 & 56737.1289 & 0.391 & 225.5 & 40.434 & \ldots & $-$25.831&  0.32   & $>$0.63 \\
         & $\pm$0.0002 & $\pm$0.0329 & $\pm$0.005 & $\pm$0.9 & $\pm$0.331 & \ldots & $\pm$0.114&  \ldots   &\ldots  \\
HIP 101941 & 1026.69 & 54442.6 & 0.223 & 211.8 & 9.975 & \ldots & $-$20.562& 0.43    & $>$0.64  \\
         & $\pm$1.37 & $\pm$7.8 & $\pm$0.012 & $\pm$2.8 & $\pm$0.111 & \ldots & $\pm$0.075&   \ldots  & \ldots \\
HIP 102320 & 21.1854 & 57143.617 & 0.416 & 322.1 & 40.879 & 41.954 & $-$21.080&  0.30   & 0.47 \\
         & $\pm$0.0002 & $\pm$0.016 & $\pm$0.003 & $\pm$0.3 & $\pm$0.141 & $\pm$0.206 & $\pm$0.049& 0.19    & 0.46 \\
HIP 104994 & 51.82128 & 55970.823 & 0.407       & 303.6     & 25.796     & 32.988     & $-$46.869&  0.85   &  0.48\\
         & $\pm$0.0015 & $\pm$0.087 & $\pm$0.003 & $\pm$0.9 & $\pm$0.090 & $\pm$0.152 & $\pm$0.062&  0.49   & 0.37 \\
HIP 110291    & 764.8 & 54423.0 & 0.563 & 3.9 & 3.888 & \ldots & $-$108.035& 0.52    & $>$0.15  \\
          & $\pm$3.5 & $\pm$4.7 & $\pm$0.028 & $\pm$4.0 & $\pm$0.219 & \ldots & $\pm$0.088& \ldots    & \ldots \\
HIP 111685 & 6177 & 54730 & 0.249 & 300.1 & 3.752 & \ldots & $-$58.378&  0.33   & $>$0.40  \\
         & $\pm$26 & $\pm$22 & $\pm$0.007 & $\pm$1.3 & $\pm$0.154 & \ldots & $\pm$0.113&  \ldots   & \ldots \\
HIP 111942 & 375.098  & 56849.88  & 0.437      & 339.5   & 7.880 & \ldots & $-$32.470   &  0.32   & $>$0.30  \\
         & $\pm$0.038 & $\pm$1.39 & $\pm$0.011 & $\pm$0.4 & $\pm$0.080 & \ldots & $\pm$0.070&  \ldots   & \ldots \\
HIP 112523 & 3255: & 55676.2 & 0.480 & 190.5 & 3.871 & \ldots & 4.154&  0.58   & $>$0.36  \\
         & $\pm$50 & $\pm$41.7 & fixed & $\pm$6.2 & $\pm$0.243 & \ldots & $\pm$0.156&  \ldots   & \ldots \\
BD+66 1664 & 3.09995 & 53785.0117 & 0.0 & 0.0 & 23.940 & \ldots & $-$22.083& 0.67    & $>$0.18  \\
         & $\pm$0.00004 & $\pm$0.0036 & fixed & fixed & $\pm$0.174 & \ldots & $\pm$0.123&  \ldots   & \ldots \\
HIP 118212 & 872.90 & 56678.80   & 0.536      & 135.4    & 8.428     & \ldots  & 0.999&  0.36   & $>$0.40  \\
         & $\pm$1.23 & $\pm$2.71 & $\pm$0.016 & $\pm$2.5 & $\pm$0.138 & \ldots & $\pm$0.131& \ldots    &\ldots  \\
\end{longtable}
}


\end{document}